\documentclass[twocolumn]{aastex631}

\newcommand{\eod}{Tau\,042021}

\begin{document}

\title{\Large JWST imaging of edge-on protoplanetary disks. I. Fully vertically mixed 10\,$\mu$m grains in the outer regions of a 1000\,au disk}

\author[0000-0002-5092-6464]{Gaspard Duch\^ene}
\affiliation{Department of Astronomy, University of California, Berkeley CA, 94720, USA}
\affiliation{Univ. Grenoble Alpes, CNRS, IPAG, 38000 Grenoble, France}

\author[0000-0002-1637-7393]{Fran\c cois M\'enard}
\affiliation{Univ. Grenoble Alpes, CNRS, IPAG, 38000 Grenoble, France}

\author[0000-0002-2805-7338]{Karl R. Stapelfeldt}
\affiliation{Jet Propulsion Laboratory, California Institute of Technology, 4800 Oak Grove Drive, Pasadena, CA 91109, USA}

\author[0000-0002-8962-448X]{Marion Villenave}
\affiliation{Jet Propulsion Laboratory, California Institute of Technology, 4800 Oak Grove Drive, Pasadena, CA 91109, USA}

\author[0000-0002-9977-8255]{Schuyler G. Wolff}
\affiliation{Department of Astronomy and Steward Observatory, University of Arizona, Tucson, AZ 85721, USA}

\author[0000-0002-3191-8151]{Marshall D. Perrin}
\affiliation{Space Telescope Science Institute, Baltimore, MD 21218, USA}

\author[0000-0001-5907-5179]{Christophe Pinte}
\affiliation{School of Physics and Astronomy, Monash University, Clayton, Vic 3800, Australia}
\affiliation{Univ. Grenoble Alpes, CNRS, IPAG, 38000 Grenoble, France}

\author[0000-0003-1451-6836]{Ryo Tazaki}
\affiliation{Univ. Grenoble Alpes, CNRS, IPAG, 38000 Grenoble, France}

\author[0000-0001-5334-5107]{Deborah L. Padgett}
\affiliation{Jet Propulsion Laboratory, California Institute of Technology, 4800 Oak Grove Drive, Pasadena, CA 91109, USA}




\begin{abstract}

Scattered light imaging of protoplanetary disks provides key insights on the geometry and dust properties in the disk surface. Here we present JWST 2--21\,$\mu$m images of a 1000\,au-radius edge-on protoplanetary disk surrounding an 0.4\,$M_\odot$ young star in Taurus, 2MASS\,J04202144+2813491. These observations represent the longest wavelengths at which a protoplanetary disk is spatially resolved in scattered light. We combine these observations with HST optical images and ALMA continuum and CO mapping. We find that the changes in the scattered light disk morphology are remarkably small across a factor of 30 in wavelength, indicating that dust in the disk surface layers is characterized by an almost gray opacity law. Using radiative transfer models, we conclude that grains up to $\gtrsim10\,\mu$m in size are fully coupled to the gas in this system, whereas grains $\gtrsim100\,\mu$m are strongly settled towards the midplane. Further analyses of these observations, and similar ones of other edge-on disks, will provide strong empirical constraints on disk dynamics and evolution and grain growth models. In addition, the 7.7 and 12.\,$\mu$m JWST images reveal an X-shaped feature located above the warm molecular layer traced by CO line emission. The highest elevations at which this feature is detectable roughly match the maximal extent of the disk in visible wavelength scattered light as well as of an unusual kinematic signature in CO. We propose that these phenomena could be related to a disk wind entraining small dust grains. 

\end{abstract}

\keywords{protoplanetary disks - planet formation - circumstellar matter - dust}


\section{Introduction} \label{sec:intro}

Gas-rich protoplanetary disks are the birth site of planetary systems. How the small ($\lesssim 1\,\mu$m) grains these disks inherit from their parent molecular cloud evolve to form km-sized planetesimals and, ultimately, rocky planets and cores of giant planets is hotly debated and multiple scenarios have been proposed \citep[see][for a recent review]{drazkowska2022}. As a first step, small grains are generally believed to grow to mm- or cm-sized pebbles by coagulation through gentle hit-and-stick interactions \citep{blum2008}. Spatially unresolved (sub)mm observations of disks surrounding T\,Tauri stars indicate that this process is largely underway by an age of $\sim1\,$Myr \citep{williams2011}. Once they have grown to large enough sizes, drag forces induced by the surrounding gas lead to their settling in the midplane and inward drift. The details of these two phenomena depend sensitively on the gas dynamics, which can be both turbulent and subject to various instabilities \citep{lesur2022}. Mapping out the degree of settling and radial drift thus informs the physics of the initial stages of planet formation and disk evolution. By modifying the local dust opacity in different regions of the disk, settling has a significant influence on even unresolved observations of protoplanetary disks \citep[e.g.,][]{dullemond2004, dalessio2006} but these effects are generally ambiguous and dependent on assumptions about the disk structure.

The advent of high resolution imaging capabilities over the past decade has provided crucial insight on dust settling. In particular, the Atacama Large Millimeter/submillimeter Array (ALMA) observations of disks have shown that the vertical extent of the (sub)mm-emitting pebbles is roughly an order of magnitude smaller than that of the gas and/or $\mu$m-sized dust that dominates scattered light \citep[e.g.,][]{pinte2016, villenave2022}. This is most readily illustrated in the case of edge-on disks (hereafter EODs), whose vertical extent is directly exposed to the observer. From a practical standpoint, these systems present the advantage that the disk blocks direct starlight, thus alleviating the need for coronagraphic techniques that are otherwise necessary. They are however severely underluminous as a result and, in general, cannot be observed with ground-based adaptive optics instruments. Space-based observations, with the Hubble Space Telescope (HST) and James Webb Space Telescope (JWST), are therefore necessary to image them at the required resolution. Achieving similarly high angular resolution from the optical to the millimeter regime is key to constraining settling in disks associated with T\,Tauri stars. Indeed, direct comparisons of HST and millimeter observations of EODs have confirmed that settling is prevalent \citep{duchene2003, villenave2020, wolff2021}. 

However, because those observational approaches are primarily sensitive to grains sizes that are several orders of magnitudes apart, they only provide constraints in two dramatically different regimes: $\mu$m-sized (and smaller) grains are fully coupled with the gas, whereas mm-sized pebbles are strongly decoupled. These are necessary constraints but it is critical to determine the behavior of intermediate-sized grains (several to a few tens of $\mu$m), which are only partially coupled to the gas, to fully disambiguate between disk and dust evolution models. Observations in the mid-infrared regime are crucial to fill this gap, as they are sensitive to such intermediate grains. In addition, thanks to the expected reduced opacity of dust at mid-infrared wavelengths, such observations should naturally probe deeper in the disk than the surface layers that are imaged with HST. Unresolved analyses of a few EODs have suggested that grains up to $\sim10\,\mu$m are present in the upper scattering layers \citep{pontoppidan2007, sturm2023a}. This has been corroborated for the few EODs that have been imaged in the 3--5\,$\mu$m range to date \citep{duchene2010, tobin2010, mccabe2011}, although only two have been resolved at 10\,$\mu$m \citep{mccabe2003, perrin2006}. The successful deployment of JWST now provides a new opportunity to image EODs in scattered light through the near- and mid-infrared ranges.

This paper focuses on the disk surrounding 2MASS\,J04202144+2813491, hereafter \eod. The object is located on the Western edge of the L1495 complex in the Taurus star forming-region, at a distance of 130\,pc \citep{galli2019}. Seeing-limited images first identified the system as an EOD, with two parallel nebulae representing the disk surfaces where stellar photons are scattered towards the observer \citep{luhman2009}. Remarkably \eod\ is one of the largest disks known in Taurus, 50\% larger than the iconic EOD HH\,30 \citep{burrows1996}. Deeper, higher-resolution HST images subsequently revealed a smooth and mostly symmetrical disk with a radius of at least 400\,au, as well as a bright collimated bipolar atomic jet \citep{duchene2014}. This indicates that the central star is actively accreting, consistent with the detection of an apparent UV excess (see Appendix\,\ref{sec:sed}). Based on near-infrared spectroscopy, \cite{luhman2009} estimated a spectral type of M1$\pm$2, suggesting an 0.4--0.8\,$M_\odot$ for the central star. \citet{andrews2013} reported the first millimeter observation of the system, with a flux density that places it in the top quartile among other Taurus members with a similar spectral type, suggesting a relatively massive disk, consistent with its large physical size. High-resolution 0.9-1.3\,mm ALMA observations were subsequently obtained by \citet{villenave2020}. These marginally resolved the disk along the vertical axis, with an apparent vertical size that is several times thinner than the separation between the two disk surfaces revealed in scattered light images. Broadly speaking, the ALMA maps support the presence of strong settling in the \eod\ disk. \citet{villenave2020} further estimate a lower limit to the disk dust mass of $8\times10^{-5}\,M_\odot$ although optical depth effects could boost this mass by up to an order of magnitude.

This paper presents the first results of an ongoing near- to mid-infrared imaging campaign of most known EODs with JWST (GO programs 2562 and 4290 in Cycles 1 and 2, co-PIs: F. M\'enard and K. R. Stapelfeldt). Specifically, we present the first 2--21\,$\mu$m JWST observations of \eod\ to further constrain the degree of settling in this disk. We also present 
CO ALMA maps to locate the warm molecular layer of the disk and to estimate the dynamical mass of the central star. Furthermore, we make use of archival GALEX, {\it Spitzer} and {\it Herschel} observations to complete the SED of the system and to probe the presence of mid-infrared emission lines. The paper is organized as follows: Section\,\ref{sec:obs} summarizes the new and archival observations used here, and the corresponding observational results are presented in Section\,\ref{sec:res}. In Section\,\ref{sec:models}, we construct several toy models, with varying degrees of settling and dust evolution to try and match the appearance of \eod\ across the wavelength range. We then discuss the implications of our findings in Section\,\ref{sec:disc}.

\section{Observations} \label{sec:obs}

\subsection{JWST} \label{subsec:obs_jwst}

We observed \eod\ with JWST as part of GO program 2562, using NIRCam and MIRI in two consecutive visits starting at 2023 January 23 UT 09:00. See Table\,\ref{tab:log} for more detail including total exposure times per filter. We used the dual channel capability of NIRCam to obtain simultaneous F200W and F444W images. With MIRI, we obtained consecutive images with the F770W, F1280W and F2100W filters. With both instruments, we used 4-point dither patterns to improve spatial sampling and mitigate detector artifacts: NIRCam used module B with the STANDARD subpixel dithering (because this target is spatially small, we did not use any large primary dither to fill in the SW inter-chip gaps), while MIRI used the BRIGHTSKY subarray and EXTENDED SOURCE dither pattern.  Exposure settings (e.g. readout pattern and number of groups per integration) were chosen to avoid saturation on target, and the number of integrations per position were set to achieve the desired total integration times for high SNR on fainter surrounding nebulosity, based on ETC predictions scaled from the HST images. 


We obtained the level-3 reduced and combined {\tt i2d} data products from the Mikulski Archive for Space Telescopes (MAST). The retrieved files were reduced with pipeline version 1.8.2 \citep{bushouse2022} and using CRDS reference file context 1041. These frames are corrected for flat-field, cosmic rays and other outliers, and optical distortion, but not for background. The background in the dither-combined NIRCam images is low and uniform and we simply subtracted the median value across the frame to remove it. This approach fails for the MIRI images due to residual gradients in the background. We therefore retrieved the individual level-2 data products from MAST, generated a background frame by median-combining the exposures at each individual location of the dither patterns, subtracted it from each frame, and manually aligned and median-combined the resulting images to produce the final images of \eod. Finally, we used a 5\arcsec\ square aperture in all JWST images to measure fluxes of 2.4, 2.6, 2.3, 1.8 and 1.9\,mJy in the F200W, F444W, F770W, F1280W and F2100W filters, respectively, with a conservative 10\%  uncertainty based on comparing these fluxes with those in larger apertures.

\begin{deluxetable}{ccccc}
 \tablecaption{Observing Log \label{tab:log}}

 \tablehead{
 \colhead{Telescope} & \colhead{Instrument} & \colhead{Filter} & \colhead{$t_{int}$ (s)} & \colhead{UT Date} }
 \startdata 
JWST & NIRCam & F200W & 1288 & 2023/01/23 \\
 & NIRCam & F444W & 1288 & 2023/01/23 \\
 & MIRI & F700W & 1052 & 2023/01/23 \\
 & MIRI & F1280W & 1284 & 2023/01/23 \\
 & MIRI & F2100W & 1326 & 2023/01/23 \\
HST & ACS & F606W & 1300 & 2011/12/10 \\
 & ACS & F814W & 840 & 2011/12/10 \\
 \enddata
\tablecomments{$t_{int}$ is the total on source integration time. Additional technical details for the JWST observations, such as detector readout settings per filter, may be found by retrieving APT program 2562 observations 1 and 2.}
\end{deluxetable}

\subsection{HST} \label{subsec:obs_hst}

We previously observed \eod\ with HST/ACS as part of Program 12514 (PI: K. R. Stapelfeldt). The observations were first presented in \citet{duchene2014} and \cite{stapelfeldt2014}. We obtained consecutive images with the F606W and F814W filters in a single orbit using {\tt CRSPLIT=2} to reject cosmic rays. The images presented here are the {\tt drz} data products available through MAST, which are corrected for flat field and optical distortion. The background level is low and uniform in both filters, and we subtracted the median value over the entire images to correct for it.

\subsection{ALMA} \label{subsec:obs_alma}

Continuum maps in ALMA bands 4, 6 and 7 were previously presented in \citet{villenave2020}, to which we refer the reader for details about observations and data reduction. Here we report the CO J $= 2 - 1$ observations associated with the already published band 6 continuum map (program 2016.1.0771.S, PI: G. Duch\^ene). The source was observed using a compact array configuration on December 5, 2016 and an extended configuration on October 21, 2016. The spectral set-up was divided into two 1.9\,GHz continuum spectral windows, of rest frequency 216.5\,GHz and 219.6\,GHz, and three spectral windows that were set-up to include the $^{12}$CO, $^{13}$CO, and C$^{18}$O J = 2 - 1 transitions at 230.538\,GHz, 220.399\,GHz and 219.56\,GHz, respectively. Here, we focus on the $^{12}$CO and $^{13}$CO observations, which were obtained with a sampling of 122\,kHz and 61\,kHz, respectively. The raw data were calibrated using the CASA pipeline version 4.7.2 \citep{casa2022}.

To maximize the dynamical range, we performed phase self-calibration on the continuum data for the compact configuration, and applied the continuum solution to the $^{12}$CO and $^{13}$CO spectral windows. We extracted the emission lines from the calibrated visibilities after subtracting continuum emission using the \texttt{uvcontsub} function in CASA. We used the \texttt{tclean} task, with a spectral resolution of 0.22\,km\,s$^{-1}$ and a Briggs robust parameter of 0.5, to produce two sets of line emission maps. For the maps combining both extended and compact configuration observations, we obtain an angular resolution of 0\farcs19$\times$0\farcs37. In addition, we also use a {\tt uvtaper} to produce a smoothed image of the $^{12}$CO channel map centered at $V-V_{sys}$= -0.3\,km\,s$^{-1}$, which achieved an angular resolution of 0\farcs74$\times$0\farcs67.

\section{Results} \label{sec:res}

\subsection{The dust disk in scattered light} \label{subsec:dustdisk}

\begin{figure}[htb]
    \centerline{\includegraphics[width=\columnwidth]{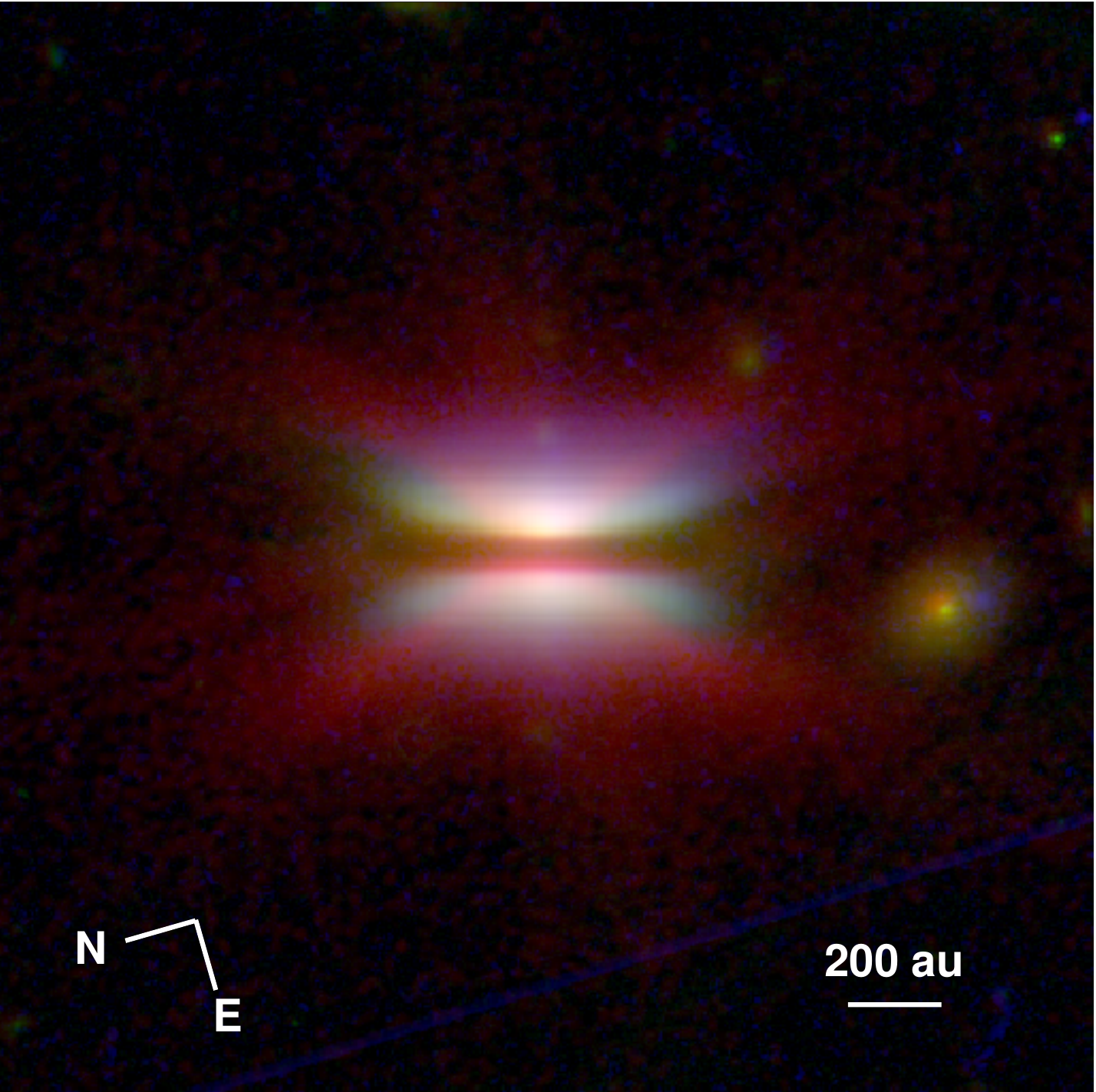}}
    \caption{HST--JWST 3-color composite image of \eod; the full field of view spans 18\arcsec. The 0.8, 2 and 7.7\,$\mu$m images are rendered (using a logarithmic stretch) in the blue, green and red channels, respectively. The orange object to the right of the edge-on disk is a background galaxy.}
    \label{fig:hst_jwst}
\end{figure}

\begin{figure*}[htb]
    \centerline{\includegraphics[width=\textwidth]{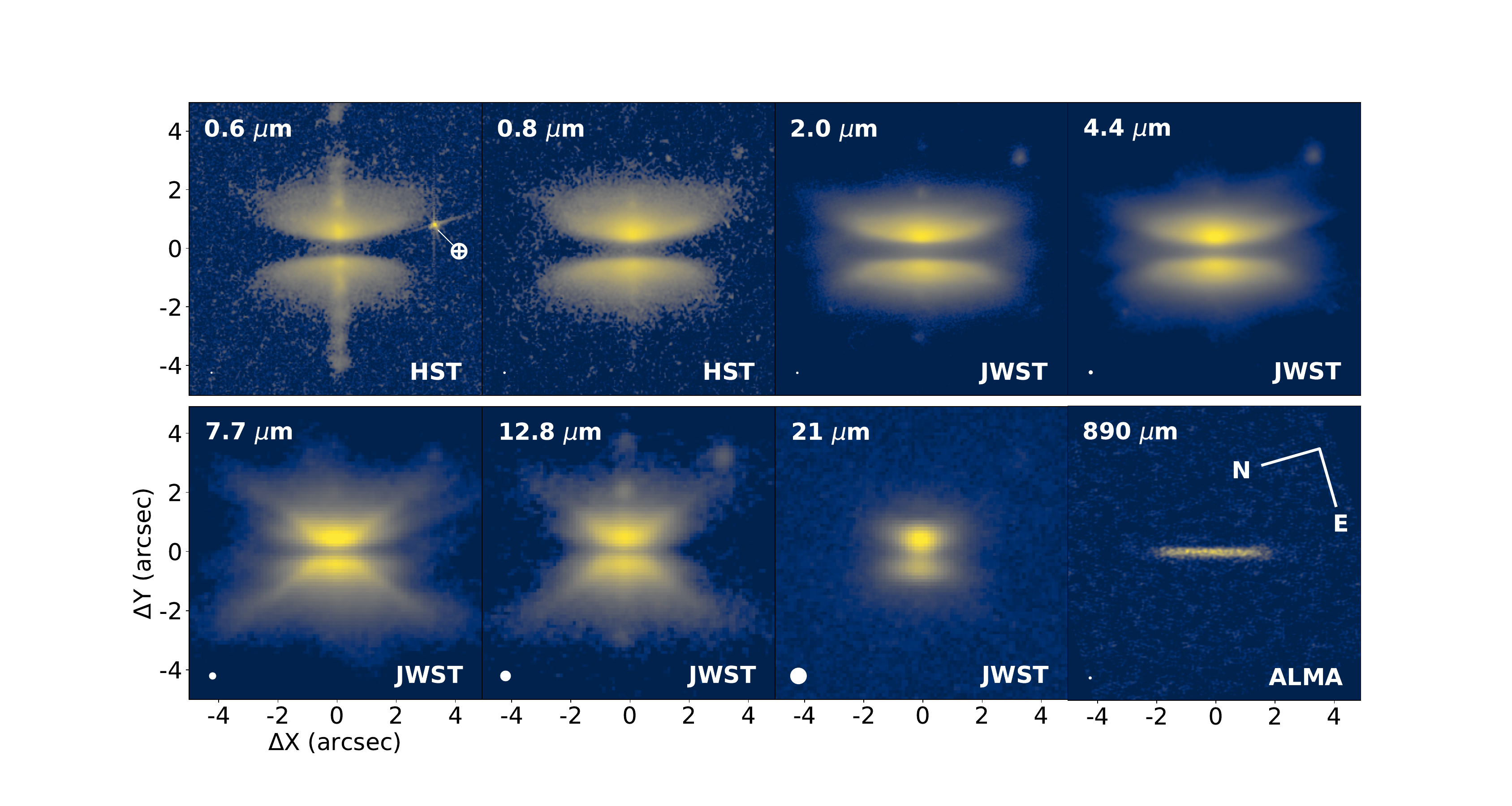}}
    \caption{Image gallery with the disk oriented horizontally. All images are shown on same angular scale and a log stretch for all panels except at 21 and 890\,$\mu$m (for which square root and linear  stretches are used, respectively). The corresponding beam is shown in the bottom left of each panel. A faint star is detected from 0.8 to 12.8\,$\mu$m to the SW of the disk. The bright spike indicated by a $\earth$ symbol in the 0.6\,$\mu$m image is an uncorrected bad pixel ``enhanced" due to Fourier shifting of the image.}
    \label{fig:gallery}
\end{figure*}

\begin{figure}[htb]
    \centerline{\includegraphics[width=\columnwidth]{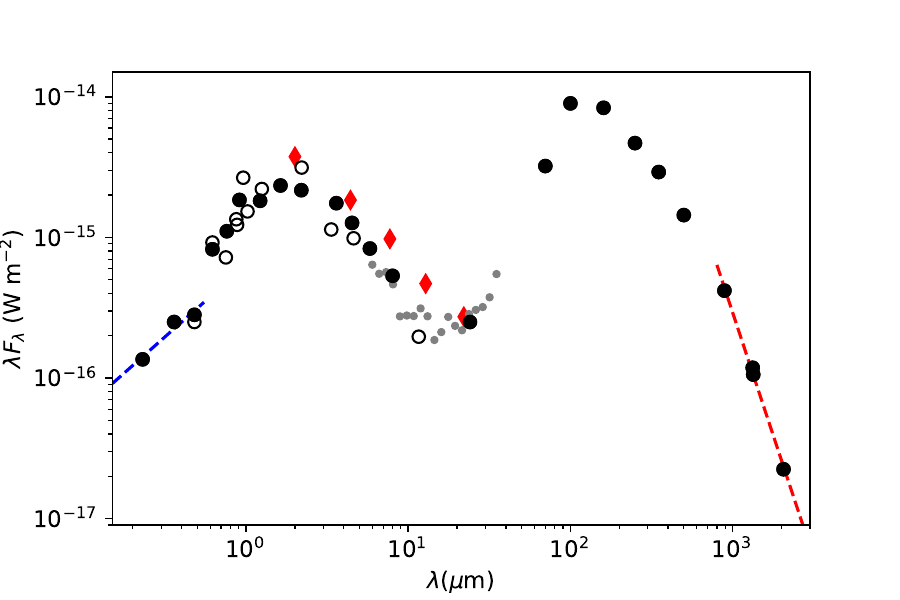}}
    \caption{SED of \eod. Closed and open circles represent the adopted and additional photometric datapoints, respectively (see Appendix\,\ref{sec:sed}).  Red diamonds indicate the fluxes measured in the JWST images presented here. Small gray circles represent the Spitzer/IRS spectrum rebinned to a $\Delta\lambda/\lambda=10\%$ sampling. The observed scatter observed from the optical to the mid-infrared is most likely due to intrinsic variability and/or changes in disk illumination. The dashed blue and red segments are power law fits to the UV/blue and (sub)mm parts of the SED, respectively. }
    \label{fig:sed}
\end{figure}

Figure\,\ref{fig:hst_jwst} presents a 3-color combination of selected HST and JWST images of \eod, whereas Figure\,\ref{fig:gallery} compares all HST and JWST images along with the ALMA band 7 continuum map. All images from the optical to the mid-infrared reveal a similar morphology, namely two nearly symmetrical nebulae bisected by a dark lane, as expected for EODs. The system's SED has its trough around 20\,$\mu$m (see Figure\,\ref{fig:sed}), and the 21\,$\mu$m image confirms that the disk is still optically thick to its own radiation out to that wavelength. In the ALMA regime, the disk is either optically thin (and emission is dominated by the denser midplane) or the two scattering surfaces are too close to each other to be fully resolved (possibly due to settling). \citet{villenave2020} showed that the intrinsic vertical extent is about 0\farcs2--0\farcs3, or 25--45\,au, with a tentative identification of separation between layers, favoring the second scenario (see below).

The overall morphology of the disk does not qualitatively change out to 5\,$\mu$m with both top and bottom nebulae showing increasing curvature as a function of distance from the symmetry axis. In the optical, faint, diffuse scattered light is detected at high elevations (up to $\approx 2$\arcsec\ from the midplane), unlike what is seen in the 2--5\,$\mu$m regime, despite comparable (or better) dynamic range in the JWST/NIRCam images. The MIRI 7.7\,$\mu$m and 12.8\,$\mu$m images reveal a markedly different structure, although the bisecting dark lane is still clearly present. On both sides of the disk a strong symmetrical X-shaped feature is present on top of a  structure that is similar to the 4.4\,$\mu$m image. This will be further explored in Section\,\ref{subsec:Xwing}. At the longest JWST wavelength, the disk is back to a simple two-nebulae structure, with a significantly reduced radial extent, and a much more centrally condensed peak, which we quantify below.

\begin{figure*}[htb]
    \centerline{\includegraphics[width=0.49\textwidth]{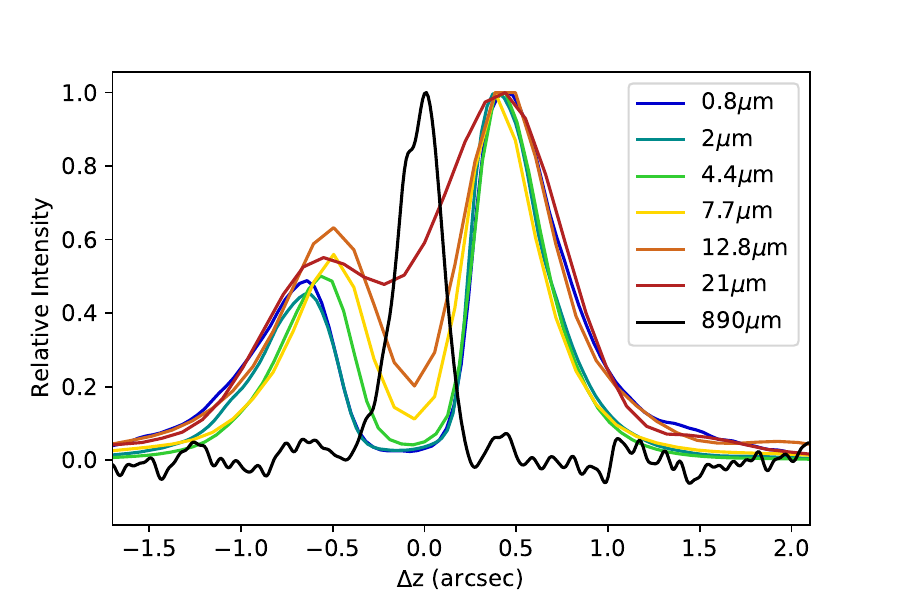}
    \includegraphics[width=0.49\textwidth]{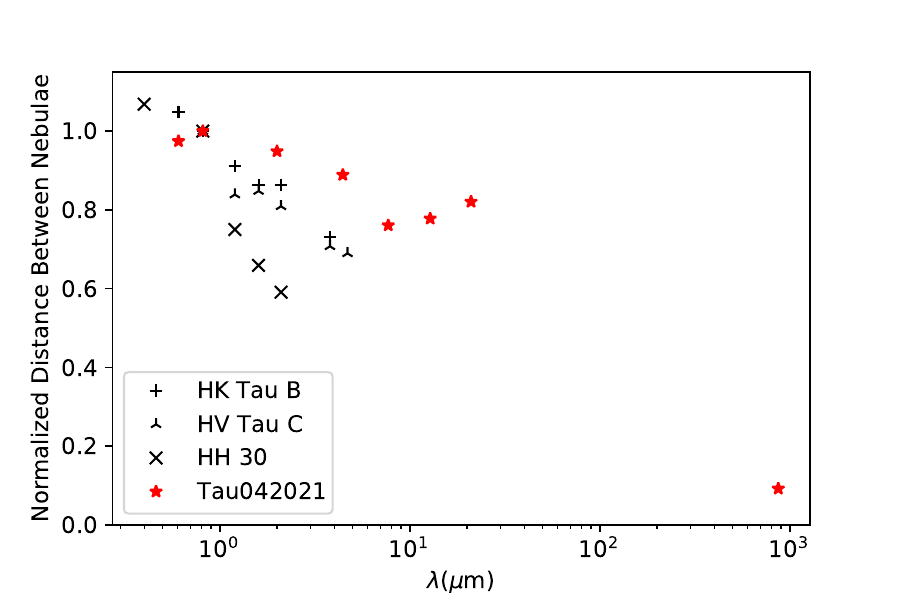}}
    \caption{{\it Left: }Vertical cut across the disk from the optical to the mid-infrared, measured within 0\farcs25 of the symmetry axis. All curves are normalized to their peak and aligned based on the location of the brighter nebula, except for the 890\,$\mu$m ALMA profile, which is aligned with the dark lane seen in the scattered light images. {\it Right: }Distance between the top and bottom nebulae, normalized to the 0.8\,$\mu$m value for comparison purposes, as a function of wavelength for \eod\ (red stars) and other EODs (black symbols). Data for other systems are from \citet{mccabe2011}, \citet{duchene2010}, and \citet{cotera2001} and \cite{watson2004} for HK\,Tau\,B, HV\,Tau\,C and HH\,30, respectively.}
    \label{fig:neb_sep}
\end{figure*}

Of interest to this study is the gradual decline in the thickness of the dark lane, measured as the distance between the peaks of the top and bottom nebulae along the disk's symmetry axis (see Figure\,\ref{fig:neb_sep}). To quantify this, we followed a similar method to that presented in \citet{villenave2020}. We first generated vertical brightness profiles, averaging within 0\farcs4--0\farcs5 windows to improve SNR, and fitted sixth to eighth order polynomials to the peak of each nebulae. We repeated the process out to 1\farcs5--3\arcsec\ (depending on filter) on either side of the symmetry axis and then fitted second order polynomials to the spine of each nebulae to evaluate the closest distance between the resulting parabolas, $d_{neb}$, as well as their peak flux ratio, FWHM and full width at 10\% of the peak (FW10\%). The latter is useful as an indication of the disk radius. Varying the polynomial degrees used in the method, we estimate uncertainties of about 0\farcs02 on $d_{neb}$. This two-step approach is more reliable than simply using vertical brightness profile along the symmetry axis for wavelengths with significant contribution from the collimated jet. As indicated in Table\,\ref{tab:disk_morphology} and shown in Figure\,\ref{fig:neb_sep}, the chromaticity of the dark lane thickness is much shallower than for other EODs \citep{watson2004, duchene2010, mccabe2011}. We defer discussion of these results to Section\,\ref{subsec:growth_settling}.

Since \eod\ is also resolved along the vertical direction in the ALMA observations \citep[][see also Figure\,\ref{fig:neb_sep}]{villenave2020}, we extracted the profile within 0\farcs5 of the symmetry axis and noticed that it presents a shoulder characteristic of two distinct components separated by a distance comparable to the projected beam size. We interpret this as an indication that the disk is still optically thick at 890\,$\mu$m, but with the two disk surfaces not fully separated. We thus fitted a double Gaussian model to this profile and estimated a distance between the two surfaces of 0\farcs11$\pm$0\farcs02, roughly ten times thinner than is observed in the optical through the mid-infrared.

\begin{deluxetable*}{cccccccc}
 \tablecaption{Disk Morphological properties \label{tab:disk_morphology}}

 \tablehead{
 \colhead{Instrument} & \colhead{Filter} & \colhead{$\lambda$} & \colhead{$d_{neb}$} & \colhead{Flux Ratio} & \colhead{FWHM$_{top}$} & \colhead{FWHM$_{bot}$} & \colhead{FW10\%$_{top}$} \\ 
  \colhead{} & \colhead{} & \colhead{($\mu$m)} & \colhead{(\arcsec)} & \colhead{} & \colhead{(\arcsec)} & \colhead{(\arcsec)} & \colhead{(\arcsec)} }
 \startdata 
HST/ACS\tablenotemark{$\dagger$} & F606W & 0.6 & 1.14 & 0.41 & 0.80 & 0.97 & 2.37 \\
HST/ACS & F814W & 0.8&  1.17 & 0.44 & 1.11 & 1.41 & 3.40 \\
JWST/NIRCam & F200W & 2.0 & 1.11 & 0.44 & 1.13 & 1.49 & 3.47 \\
JWST/NIRCam & F444W & 4.4 & 1.04 & 0.49 & 0.99 & 1.25 & 2.69 \\
JWST/MIRI\tablenotemark{$^\#$} & F770W & 7.7 & 0.89 & 0.56 & 1.10 & 1.62 & 2.50 \\
JWST/MIRI\tablenotemark{$\dagger ^\#$} & F1280W & 12.8 & 0.91 & 0.61 & 1.21 & 1.63 & 2.75 \\
JWST/MIRI & F2100W & 21 & 0.96 & 0.55 & 1.18 & 1.59 & 2.85 \\
 \enddata
\tablecomments{The distance between the apexes of the top and bottom nebulae, $d_{neb}$, is estimated based on fitting a parabola to each spine (see Section\,\ref{subsec:dustdisk}). Their flux ratio is measured along the vertical profile at the location corresponding to this smallest separation. The FWHM of both nebulae and full width at 10\% of the top nebula are estimated from the brightness profile along their respective spines.}
\tablenotetext{\dagger}{Dataset with significant contamination by line emission along the collimated jet.}
\tablenotetext{\#}{ Dataset with significant contamination by the X feature.}
\end{deluxetable*}

In the HST images, the disk presents a significant lateral asymmetry in its outer regions (away from the axis of symmetry), with the top and bottom nebulae extending further in opposite directions from one another. In contrast, in the JWST images, such an asymmetry is strongly suppressed across all wavelengths.There is therefore temporal variability in the illumination from the central star, in line with the findings from \citet{luhman2009}. Unfortunately, there is currently not enough temporal coverage to determine whether this variability is periodic.

The brightness ratio between top and bottom nebula in an EOD is related to the system's inclination for symmetric disk, via the dust scattering phase function and extinction through the disk \citep{watson2007}. The brightness ratio between the two nebulae is $\approx0.5$ (see Table\,\ref{tab:disk_morphology}), suggesting a small deviation from the $i=90$\degr\ configuration. We note, however, that this ratio has been found to vary and occasionally approach unity \citep{luhman2009}, which precludes a definitive estimate. In addition, we find a mild increase from the optical to the mid-infrared, from $\approx0.45$ to $\approx0.6$. By comparison, this flux ratio was found to strongly increase and decrease in the cases of HK\,Tau\,B and HV\,Tau\,C, respectively \citep{duchene2010, mccabe2011}.

The surface brightness profile along the spine of each nebulae is also (indirectly) driven by the scattering phase function since it probes scattering angles that range from $\approx0$\degr\ to $\approx90$\degr\ (along the major axis and at the tips of the each nebula, respectively). As indicated in Table\,\ref{tab:disk_morphology}, both the top and bottom nebulae have essentially achromatic FWHM, with the fainter (bottom) nebula being broader than the brighter (top) one. This results from an enhanced role of multiple scattering for photons escaping through the bottom layer. This is in line with the behavior observed for HV\,Tau\,C \citep{duchene2010}. Instead, in HK\,Tau\,B, the images become increasingly peakier at longer wavelengths \citep{mccabe2011}, suggesting that there is diversity in the dust scattering properties of EODs. In the mid-infrared, the \eod\ disk appears to be somewhat more compact as measured by FW10\%$_{top}$, with a decrease of 20--30\% relative to the optical and near-infrared.

Finally, we note the presence of a background galaxy (clearly identified as such in the JWST 2\,$\mu$m image, see Figure\,\ref{fig:hst_jwst}) at a projected distance of 6\farcs6, slightly interior to the gaseous disk outer edge (see Section\,\ref{subsec:gasdisk}). The galaxy is faintly detected in the HST F814W image but not in the F606W. We thus conclude that the disk must be optically thin at least at $\lambda \gtrsim 0.8\,\mu$m at this location, revealing a low column density of dust in these outer regions. Furthermore,  to estimate the proper motion of \eod\ between the HST and JWST observing epochs, we evaluated the centroid position of the galaxy in the 0.8\,$\mu$m and 2\,$\mu$m images, respectively, and measured the offset from the location of peak surface brightness in the disk (center of the brighter nebula). This yields a proper motion for \eod\ of about 28\,mas/yr South and 11\,mas/yr East, with an uncertainty of about 4\,mas/yr based on an estimated $\approx$1\,pixel uncertainty in the disk position. This result is consistent with membership to the L1495 cloud \citep{galli2019}. Interestingly, the projected separation between \eod\ and the background galaxy is declining so that photometric monitoring of the galaxy could be used as a probe of dust extinction in the disk over the next few decades.

\subsection{The collimated jet} \label{subsec:jet}

\begin{figure}[htb]
    \centerline{\includegraphics[width=\columnwidth]{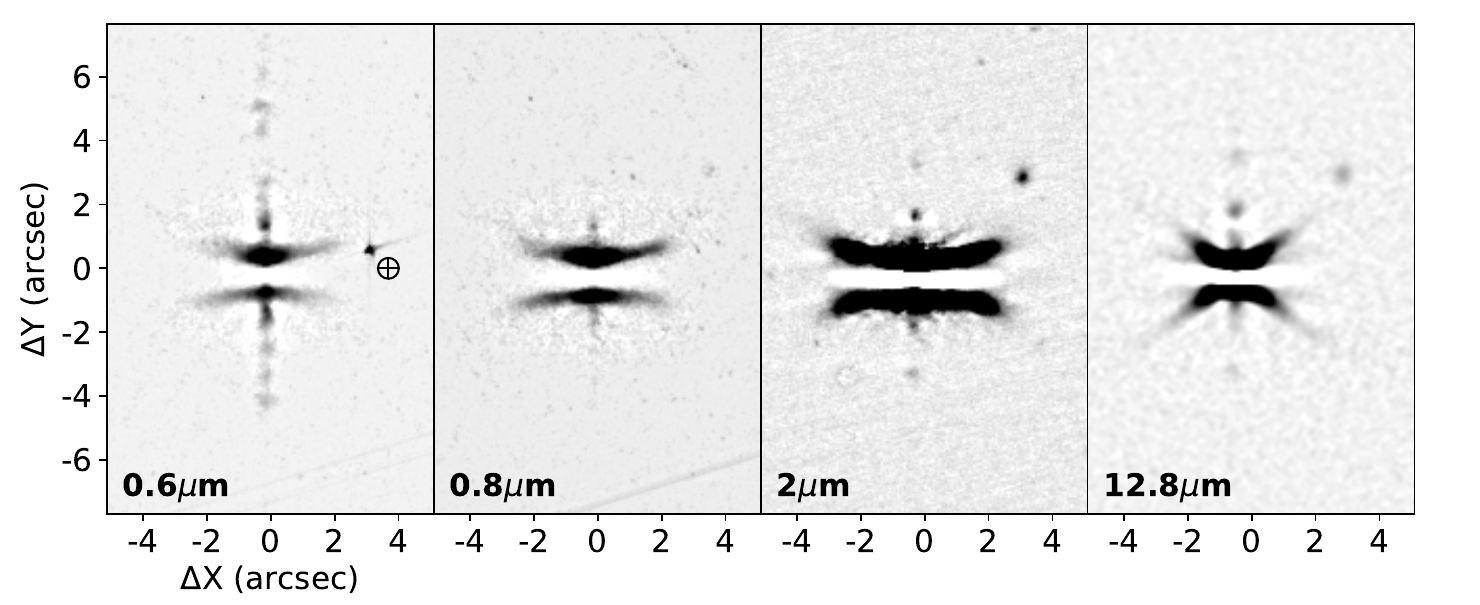}}
    \caption{HST and JWST images of \eod\ after application of a 1\arcsec\ box median filter. The images are rendered with a hard logarithmic stretch to highlight the jet component of the system, leading to saturation in the scattered light disk and X-shaped feature areas. }
    \label{fig:jet}
\end{figure}

Jet emission in the form of discrete knots is clearly detected in the HST and some of the JWST images. To highlight these features, we applied a 1\arcsec\ box median filter to these images; the resulting high-passed images are shown in Figure\,\ref{fig:jet}. A series of prominent emission knots can be traced up to $\approx6$\arcsec\ in the HST 0.6\,$\mu$m image. Jet emission in this filter is typically produced by a combination of H$\alpha$ and several nearby forbidden lines, in particular [\ion{O}{1}], [\ion{N}{2}] and [\ion{S}{2}] \citep{ray2007}. Fainter emission is also identified in the same knots in the 0.8\,$\mu$m image. We speculate that these are due to [\ion{O}{2}], [\ion{S}{3}] or [\ion{Fe}{2}] forbidden lines, which have been seen in some jet-driving T\,Tauri stars with near-edge-on viewing geometry \citep{bacciotti2011, whelan2014}.

\begin{figure}[htb]
    \centerline{\includegraphics[width=\columnwidth]{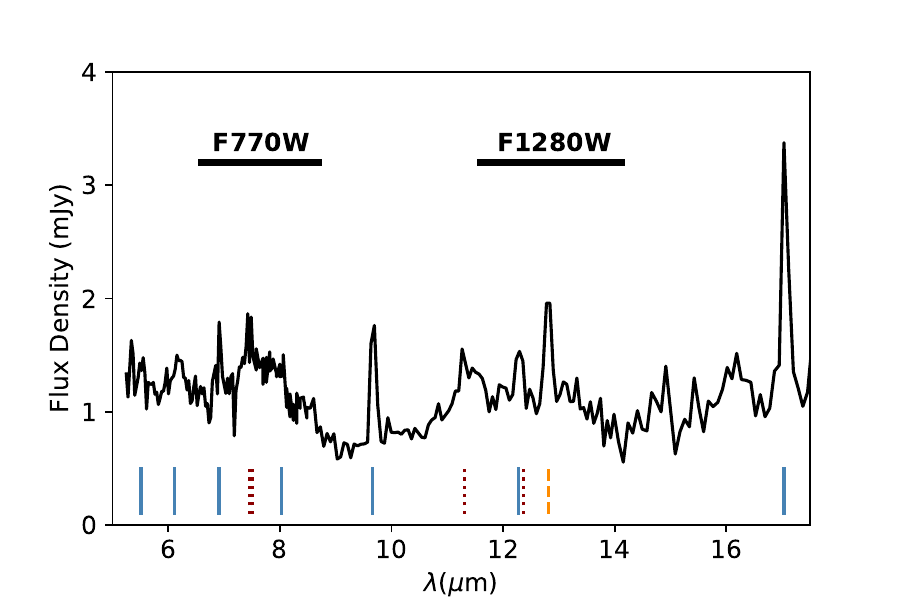}}
    \caption{{\it Spitzer} IRS spectrum of \eod\ with H$_2$, H and [\ion{Ne}{2}] lines indicated by the solid blue, dotted red and dashed orange segments, respectively.}
    \label{fig:irs_spec}
\end{figure}

Several knots are similarly identified in the JWST 2\,$\mu$m and 12.8\,$\mu$m images. In the infrared, H$_2$ and [\ion{Ne}{2}] are among the dominant outflow-related emission line \citep{lahuis2007}. To confirm that these lines are the likely tracers of the jet knots in the \eod\ system, we obtained the {\it Spitzer}/IRS spectrum of \eod\ from the Combined Atlas of Sources with {\it Spitzer} IRS Spectra \citep[CASSIS,\footnote{https://cassis.sirtf.com/atlas/welcome.shtml}][]{lebouteiller2011}. Although the source is faint ($\approx1$\,mJy), the detection is solid and the spectrum is of good quality across the entire 5--35\,$\mu$m range. Of particular interest for this study is the $\lambda \leq 15\,\mu$m regime, which contains the F770W and F1280W bandpasses in which the X-shaped feature is detected (see Section\,\ref{subsec:Xwing}). As shown in Figure\,\ref{fig:irs_spec}, there are multiple emission lines present in the spectrum, with the most prominent being the 12.8\,$\mu$m [\ion{Ne}{2}] line and the H$_2$ S(1)--S(7) series \citep[some of these lines were already reported in][]{sturm2023b}. The former is commonly associated with disk winds and/or outflows in young stars \citep[e.g.,][]{gudel2010, pascucci2020} whereas the latter is generally thought to be due to UV excitation in the uppermost layers of disks and in photodissociative winds \citep[e.g.,][]{nomura2005, lahuis2007}. We also find tentative detections of the H\,I 6--5, 8--6, 9--7 and 7--6 lines (7.46, 7.50, 11.31 and 12.37\,$\mu$m, respectively), which are thought to be driven by accretion on the central star \citep{rigliaco2015}. 

\subsection{A new X-shaped feature} \label{subsec:Xwing}

\begin{figure}[htb]
    \centerline{\includegraphics[width=\columnwidth]{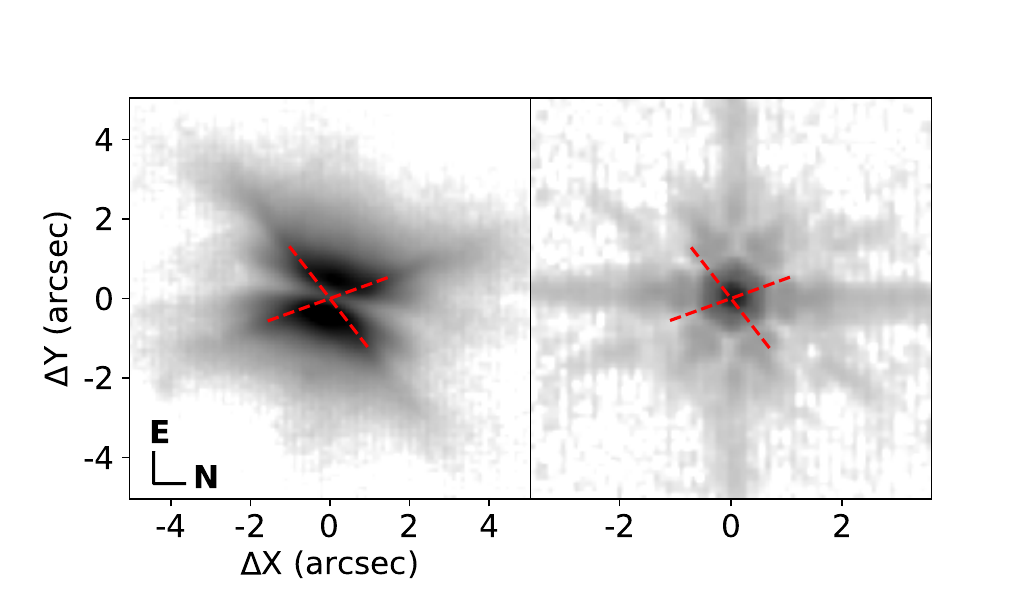}}
    \caption{MIRI 7.7\,$\mu$m images of \eod and a bright star in the same frame (left and right panels, respectively). Both images are shown in the native detector orientation and with a hard log stretch to highlight the X-shaped feature and the low-level structure of the PSF. Notice the smaller field-of-view of the right panel. To guide the eye, red dashed segments are drawn at 36\degr\ on either side of the disk midplane and replicated in the right panel.}
    \label{fig:Xwing_psf}
\end{figure}

The 7.7\,$\mu$m and 12.8\,$\mu$m images reveal a prominent X-shaped feature with an essentially linear shape that is absent in all other images of the system. Such straight features are often a consequence of telescope diffraction and/or detector properties, and JWST/MIRI is known for displaying such structures in some filters \citep[e.g.,][]{gaspar2021}. To assess whether this feature is instrumental or astrophysical, we compare the F770W image of \eod\ with that of a bright point source (2MASS\,J04202192+2813206) observed in the same frame in Figure\,\ref{fig:Xwing_psf}. None of the diffraction features of the hexagonal primary mirror line up with the X-shaped feature, nor does the so-called ``cross artefact" that runs along the detector columns and rows. The X-shaped feature associated with \eod\ is therefore astrophysical in nature. Given the bandpasses of the F770W and F1280W filters, we interpret the emission in this feature as due to photodissociated H$_2$, shock-excited [Ne\,II] and/or thermal emission from out-of-equilibrium very small grains, either at the disk surface or at the base of disk wind. This can be confirmed with integral field unit JWST observations as these have different spectral characteristics. 

\begin{figure}[htb]
    \centerline{\includegraphics[width=\columnwidth]{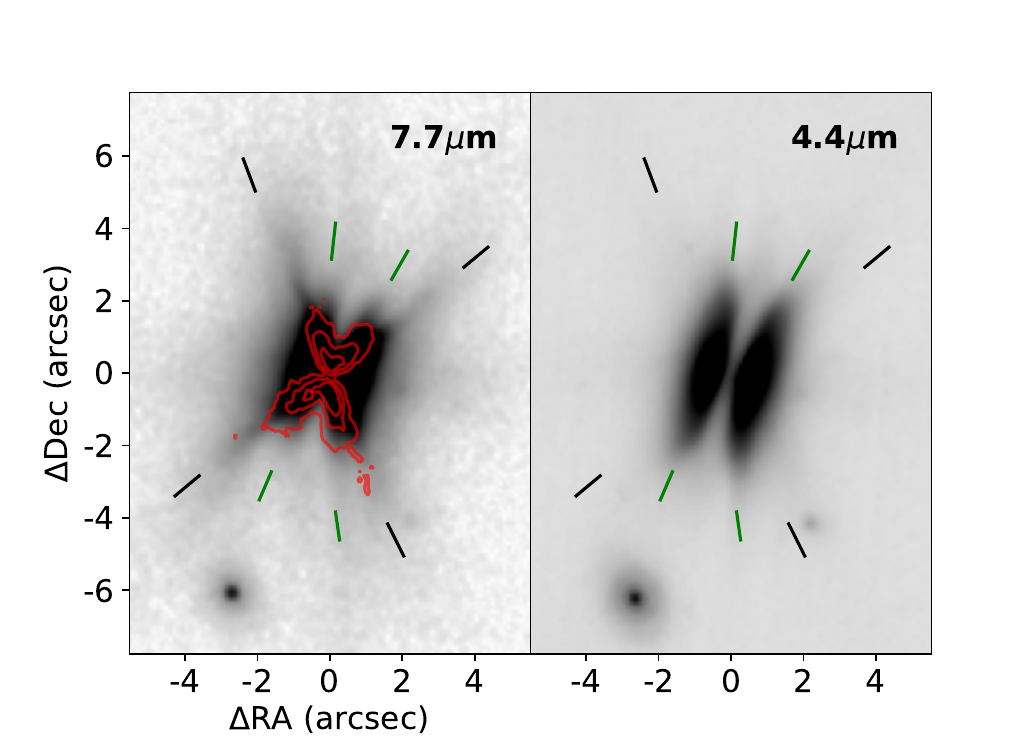}}
    \caption{MIRI 7.7\,$\mu$m and NIRCam 4.4\,$\mu$m images of \eod, with moment zero $^{12}$CO contours overplotted in the left panel. Black and green segments indicate the arms of the X-shaped feature and the disk surface orientation as seen in the F770W, respectively, and are reproduced identically in both panels as a visual guide. }
    \label{fig:Xwing_co}
\end{figure}

Figure\,\ref{fig:Xwing_co} shows the 7.7\,$\mu$m image of \eod\ with a stretch that highlights the faint outer substructures. In addition to the X-shaped feature, fainter whiskers are seen extending at similar position angles as the disk surface traced in the 4.4\,$\mu$m image. To better characterize the location of this feature, we also compare the 7.7\,$\mu$m image to the molecular warm layer, as determined by the moment zero map of the $^{12}$CO emission (see Section\,\ref{subsec:gasdisk}). It appears that the warm molecular layer is located between the 5--8\,$\mu$m scattered light disk ``surface" and the X-shaped feature itself.

The ``arms" of the X-shaped feature appear remarkably linear and are well represented by symmetrical straight lines inclined $\approx36$\degr\ from the disk midplane. The location where the two lines intersect aligns well with the jet features seen in the 12.8\,$\mu$m image (see Section\,\ref{subsec:jet}) and we therefore assume that it is the location of the central star. We further note that the arms of the feature are resolved along their minor axis, with a FWHM of $\sim$0\farcs6 in the 7.7\,$\mu$m image, i.e., more than twice the instrumental resolution. It is likely that the X-shaped feature is the limb-brightened edge of a biconical structure, whose nature will be discussed in Section\,\ref{subsec:wind_pahs}. This feature can be traced up to an elevation of 1\farcs5--1\farcs75 on either side of the midplane, almost as high as the most vertically extended scattered light detected in the HST optical images.

\subsection{The gaseous disk} \label{subsec:gasdisk}

The ALMA spectral cubes show a strong detection of the $^{12}$CO and $^{13}$CO emission, which enables us to 1) estimate the dynamical mass of the central star via its (near) Keplerian motion, and 2) probe the full radial extent of the disk. 

\begin{figure}[htb]
    \centerline{\includegraphics[width=0.9\columnwidth]{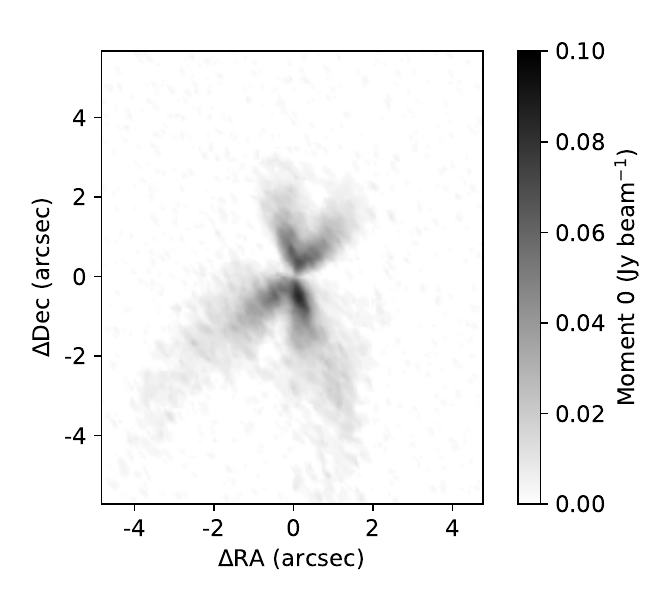}}
    \centerline{\includegraphics[width=0.9\columnwidth]{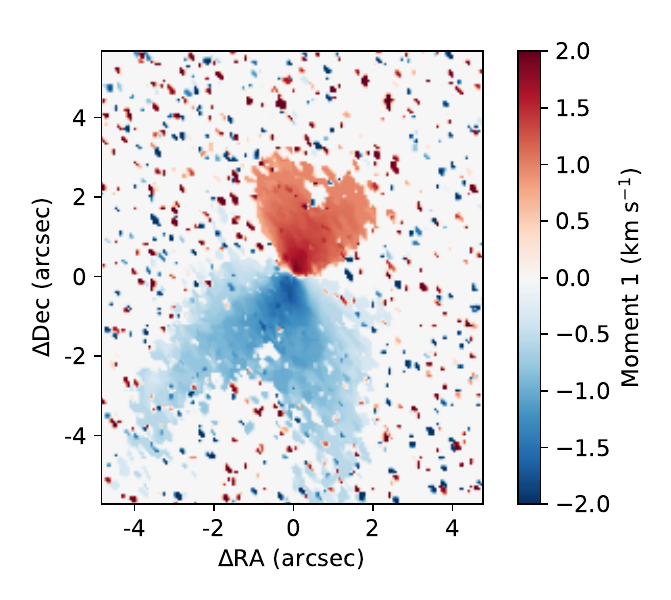}}
    \caption{Moment 0 and 1 maps of the $^{12}$CO emission from \eod\ (top and bottom, respectively). A 3$\sigma$ clipping was applied in each datacube prior to computing the moment maps.}
    \label{fig:momentmaps}
\end{figure}

Inspection of the spectral cubes reveals that the northern side of the disk is rotating away from the observer. By comparing the blue- and red-shifted channels, we estimate that $V_{sys} = 7.4$\,km\,s$^{-1}$ consistent with the $\approx7.0\pm0.5$\,km\,s$^{-1}$ range observed in the L1495 cloud \citep{galli2019}. This is further confirmation that \eod\ is physically associated with that cloud within the Taurus complex. We also note that the $^{12}$CO disk emission is heavily suppressed in the $V_{LSR} = 7$--8\,km\,s$^{-1}$ range, consistent with cloud emission \citep{goldsmith2008} that is filtered by the interferometer. Apart from this, the moment maps (Figure\,\ref{fig:momentmaps}) show the typical morphology of EODs \citep{dutrey2017, louvet2018}, with the two warm molecular layers separated by a cold midplane where CO freezes out on dust grains.

\begin{figure}[htb]
    \centerline{\includegraphics[width=\columnwidth]{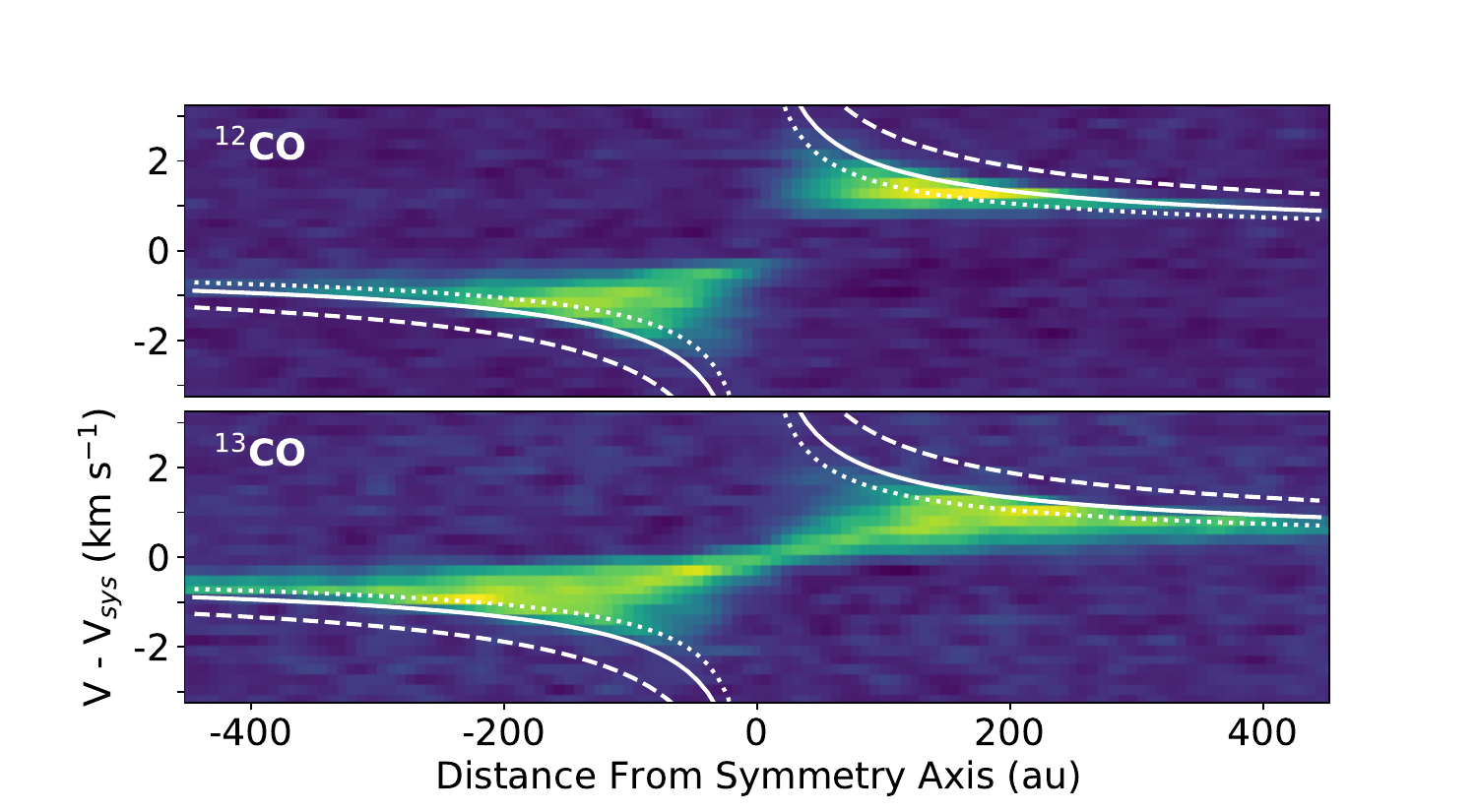}}
    \caption{$^{12}$CO and $^{13}$CO position-velocity diagrams (top
 and bottom, respectively). The dotted, solid and dashed curves correspond to dynamical masses of 0.25, 0.4 and 0.8\,$M_\odot$, respectively. The lack of low-velocity $^{12}$CO emission is due to interferometric filtering of fore/background cloud emission.}
    \label{fig:PVdiag}
\end{figure}

We then constructed the $^{12}$CO and $^{13}$CO position-velocity diagrams and superimposed representative Keplerian curves (Figure\,\ref{fig:PVdiag}). A coarse exploration suggests a stellar mass in the 0.3--0.4\,$M_\odot$ range. We note that this estimate is consistent with the lower end of the range derived from the system's spectral type while the higher end of that range is firmly excluded by observations \citep{luhman2009}. Conversely, the 0.25\,$M_\odot$ estimated by \citet[][after scaling to the 130\,pc distance used here]{Simon2019} is too low. We note that the best-fitting Keplerian model derived by \cite{Simon2019} present significant residuals, which may be due to a systematic bias at high inclination induced by the methodology employed by these authors or to the presence of non-Keplerian motion (see below). While obtaining a precise estimate of the stellar mass is beyond the scope of this study since it does not directly affect our analysis, we conclude that $M_\star \approx 0.4\,M_\odot$. 

\begin{figure}[htb]
    \centerline{\includegraphics[width=\columnwidth]{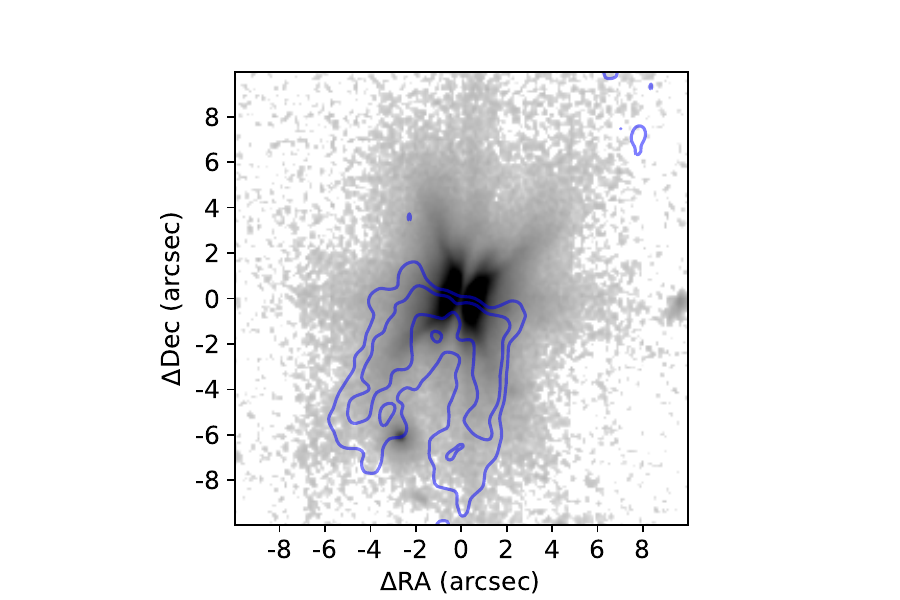}}
    \caption{$^{12}$CO channel map centered at $V - V_{sys} = -0.3$\,km\,s$^{-1}$ (contours at 20 and 50\% of the peak brightness) and using (u,v) tapering to a 0\farcs7 circular beam to improve sensitivity in the outer disk, superimposed over the MIRI F770W image, stretched to highlight the direction and extent of the X-shaped feature.}
    \label{fig:COsinglechannel}
\end{figure}

The moment maps are well defined up to a radius of about 2\arcsec, significantly less than the HST or JWST radial extent but marginally larger than the submillimeter continuum \citep{villenave2020}. 
However, faint $^{12}$CO emission is detected out to much larger distances and is best mapped in a few individual channels. On the blueshifted (South) side of the disk, the channel centered at $V - V_{sys} = -0.3$\,km\,s$^{-1}$ reveals detectable emission in the disk upper layers out to a radius of at least $\approx8$\arcsec ($\approx1000$\,au, see Figure\,\ref{fig:COsinglechannel}). The fact that the CO emission is most extended at such a low velocity (relative to the systemic velocity) is highly unusual among EODs \citep[e.g.,][]{guilloteau2016, louvet2018, teague2020} and suggests non-Keplerian motion. Unfortunately, the corresponding red-shifted channel is devoid of emission due to filtered-out cloud emission. Nonetheless, this measured size is twice as much as in the HST and JWST images, and $\approx3.5$ times more than in the submillimeter continuum emission. This is in line with the typical ratio between gas and dust (submillimeter) component of protoplanetary disks, both edge-on and at lower inclinations \citep{ansdell2018, flores2021}. This channel map shows a notably boxy pattern, essentially flat-topped, suggesting that the warm molecular layer extends over a wide range of radii at a fixed elevation of $\sim325$\,au, a height that matches the end points of the X-shaped feature discussed in Section\,\ref{subsec:Xwing}. 

\section{Modeling} \label{sec:models}

\subsection{Objectives and modeling setup} \label{subsec:models_setup}

With the data presented here, the \eod\ disk is now extremely well characterized, with scattered light images spanning a factor of 30 in wavelengths, several thermal emission maps in the (sub)millimeter regime, CO (2--1) spectral cubes, and a well sampled SED that includes a {\it Spitzer}/IRS spectrum. While simultaneously modeling all of these observables is the principled methodology to characterizing the disk structure in great detail, it is beyond the scope of this study. Instead, we adopt an exploratory approach that focuses on two key physical aspects, the grain size distribution (indicative of grain growth) and the degree of settling in the disk. To probe these phenomena in an efficient manner, we consider the thickness of the dark lane that separate the two scattered light surfaces ($d_{neb}$) as a direct, albeit reductive, observable tracer of dust opacity and disk vertical structure. All model calculations are performed with the MCFOST radiative transfer code \citep{pinte2006, pinte2009}.

Concretely, we adopted the following methodology. We first fix the stellar properties based on our current knowledge of the system. Specifically, we adopt a stellar mass of 0.4\,$M_\odot$ (Section\,\ref{subsec:gasdisk}) and an effective temperature of 3700\,K based on the system's spectral type and adopting the conversion law from \citet{luhman2003}. We further set the stellar luminosity to $L_\star = 0.85\,L_\odot$ based on the median luminosity of Taurus members with the same spectral type \citep{luhman2009}. Our model also includes a UV excess component that is evident in the short wavelength end of the system's SED. Fitting a power law to the datapoints at $\lambda < 0.5\,\mu$m, we find that this excess is characterized by a $F_\nu \propto \nu^{-2}$ shape and a total excess of $F_{\mathrm{UV}} / F_\star = 4$\% (see Figure\,\ref{fig:sed}). Finally, we include no foreground extinction, consistent with the GALEX detection of \eod.

To describe the density structure of the gaseous component, we adopt the commonly used tapered edge surface density profile \citep[see, e.g.,][]{williams2011, andrews2015}:
$$\ \Sigma(r)\ \propto \ r^{-\gamma}\ \exp\left[ - \left(\frac{r}{r_c}\right)^{2 - \gamma_{\exp}}\right],$$
where we set $\gamma = \gamma_{\exp} = 1$ and $r_c = 300$\,au, coupled with a hard outer cutoff at $r_{out} = 400$\,au. These values were selected based on an initial manual exploration of the parameter space as they produced an acceptable extent along the disk major axis. We assumed a Gaussian vertical density profile, appropriate for a vertically isothermal disk, characterized by a scale height of $h_0 =11$\,au at $r_0 = 100$\,au and a flaring exponent of $\beta = 1.125$ so that $h(r) = h_0 (\frac{r}{r_0})^\beta$. The value for $h_0$ was computed from Eq.\,6 in \citet{andrews2015}, using the midplane temperature of the same initial models. The vertically isothermal assumption is incorrect as the disk surface is super-heated but the resulting difference in vertical density profile is most noticeable in the optically thin highest regions of the disk that do not contribute significantly to scattered light images. Besides, the assumption is only employed to set the vertical density structure; the temperature is subsequently computed locally during the radiative transfer calculation.

 Finally, we selected an inclination of $i=88$\degr, which generally leads to peak flux ratios of $\approx2$ between the two disk surfaces, consistent with the observations presented here. While model scattered light images are dependent on all of these parameters, our analysis on the wavelength dependence of the dark lane thickness is not strongly affected as it is a chromatic effect. Thus, fixing these parameters reduces dramatically the dimensionality of the problem while keeping the focus on the topic of interest here, namely the dust properties and especially their spatially-dependent size distribution. Nonetheless, a more thorough exploration of the disk structural parameters will be required in future modeling of the source. As described below, we explore models in which dust particles of all sizes are fully mixed with the gas, as well as some models in which larger grains are settled closer to the midplane.

\begin{figure}[htb]
    \centerline{\includegraphics[width=0.49\textwidth]{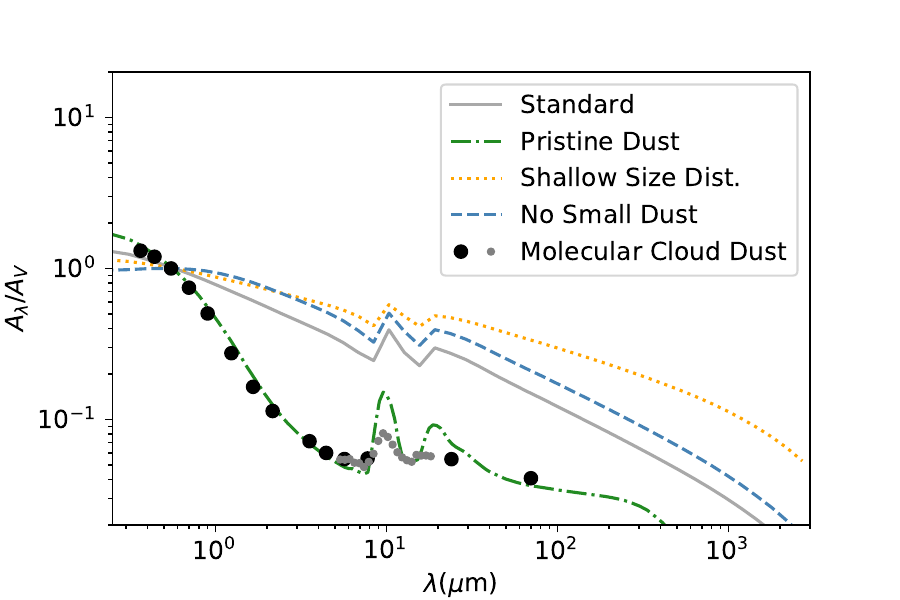}}
    \caption{Extinction law of dust models used in this study. The black and gray symbols represent the extinction law observed in molecular clouds (see Appendix\,\ref{sec:dust_molcloud}).}
    \label{fig:dust_prop}
\end{figure}

In the absence of direct quantitative constraints on the dust composition in the \eod\ disk, we adopted the DIANA mixture, namely a 60/15/25 (volume fraction) mixture of silicates, amorphous carbon and vacuum \citep{woitke2016}. Instead of the standard Mie theory, applicable for compact spherical grains, this model uses the Distribution of Hollow Spheres method \citep[DHS, ][]{min2016}, which provides absorption and scattering properties that mimic aggregate dust particles. In combination with the selected porosity, this treatment of dust scattering produces stronger forward scattering. In turn, this leads to more centrally peaked images of EODs (since the central part of the top disk surface is seen at very small scattering angles), which matches better with our observations of \eod. 

Regarding the grain size distribution, we first consider a ``standard" model that is a $\mathrm{d}n(a) \propto a^{-p}\,\mathrm{d}a$ power law with $p=3.5$ ranging from $a_\mathrm{min} = 0.03\,\mu$m to $a_\mathrm{max} = 1$\,mm, although we will also explore deviations from these default values (see Table\,\ref{tab:model_res}). Furthermore, we consider a ``pristine" model, in which we built a grain size distribution that produces an extinction law consistent with that observed in molecular clouds out to $\lambda = 70\,\mu$m, albeit with moderately too strong mid-infrared silicate features (see Figure\,\ref{fig:dust_prop}). We refer the reader to Appendix\,\ref{sec:dust_molcloud} for the details of this model. We stress that this model should not be considered as physically realistic but it achieves the desired opacity law, which is the quantity we wish to probe in the \eod\ disk. 

Since empirical evidence for dust settling of large (millimeter-emitting) dust in protoplanetary disks is now strong \citep[e.g.,][]{villenave2020}, we incorporate a simple settling prescription to explore its qualitative influence on the JWST images of \eod. Rather than adopting a specific physical model, which depends on poorly constrained ingredients such as the level of turbulence in a disk and the microphysics of grain-gas interactions, we adopt a simple analytical prescription, whereby grains with $a > a_\mathrm{mix}$ have a reduced scale height: $h(a) \propto (a/a_\mathrm{mix})^{\eta_\mathrm{settl}}$, with $\eta_\mathrm{settl} < 0$. Expectations from hydrodynamical and magneto-hydrodynamical turbulent settling suggest $\eta_\mathrm{settl} \gtrsim -0.5$ \citep[e.g.,][]{dubrulle1995, fromang2009}, while ALMA observations indicate that $h(a \gtrsim 100\mu\mathrm{m}) \approx 1$\,au \citep{pinte2016, villenave2020}.

For each model (i.e., different size distribution and settling parameters), we then adjusted the disk mass to match the observed dark lane thickness at 2\,$\mu$m. We selected the JWST 2\,$\mu$m disk image for this as it has the best contrast/resolution combination in our datasets. Having set this last parameter, we then compute images at other wavelengths from the optical to the sub-millimeter as well as the complete SED, and determine whether it is an acceptable fit to the observations (see Table\,\ref{tab:model_res}).

\subsection{Modeling results} \label{subsec:model_results}

\begin{table*}
\begin{center}
\caption{Summary of Model Exploration \label{tab:model_res}}
\begin{tabular}{cccccccccc}
\hline
Model & \multicolumn{3}{c}{Grain Size Distribution} & \multicolumn{2}{c}{Settling} & Total & Dark Lane & 890\,$\mu$m & SED \\
  & $p$ & $a_\mathrm{min}$ & $a_\mathrm{max}$ & $\eta_\mathrm{settl}$ & $a_\mathrm{mix}$ & Dust Mass & Thickness & Map &  \\
 &  & ($\mu$m) & ($\mu$m) &  & ($\mu$m) & ($10^{-4}\,M_\odot$) &  &  &   \\
\hline
Standard & 3.5 & 0.03 & 1000 & \multicolumn{2}{c}{None} & 4.9 & (\checkmark) & $\times$ & (\checkmark) \\
Pristine Dust & \multicolumn{3}{c}{see Appendix\,\ref{sec:dust_molcloud}} & \multicolumn{2}{c}{None} & 2.8 & $\times$ & $\times$ & $\times$ \\
Shallow Size Distribution & 3.25 & 0.03 & 1000 & \multicolumn{2}{c}{None} & 13 & \checkmark & $\times$ & (\checkmark) \\
No Small Dust & 3.5 & 0.3 & 1000 & \multicolumn{2}{c}{None} & 5.1 & \checkmark & $\times$ & $\times$ \\
Settling ($a_\mathrm{mix} = 100\,\mu$m) & 3.5 & 0.03 & 1000 & 0.5 & 100 & 4.5 & (\checkmark) & $\times$ & (\checkmark) \\
Settling ($a_\mathrm{mix} = 10\,\mu$m) & 3.5 & 0.03 & 1000 & 0.5 & 10 & 4.9 & (\checkmark) & (\checkmark) & (\checkmark) \\
Settling ($a_\mathrm{mix} = 1\,\mu$m) & 3.5 & 0.03 & 1000 & 0.5 & 1 & 7.0 & $\times$ & (\checkmark) & $\times$ \\
\hline
\end{tabular}
\end{center}
\tablecomments{The symbols in the last three columns describe the degree to which a given model matches observations: a \checkmark\ symbol represents an acceptable quantitative match, a (\checkmark) indicates a small but significant deviation, while a $\times$ symbol points to a large shortcoming of the model.}
\end{table*}

\begin{figure*}[htb]
    \centerline{\includegraphics[width=\textwidth]{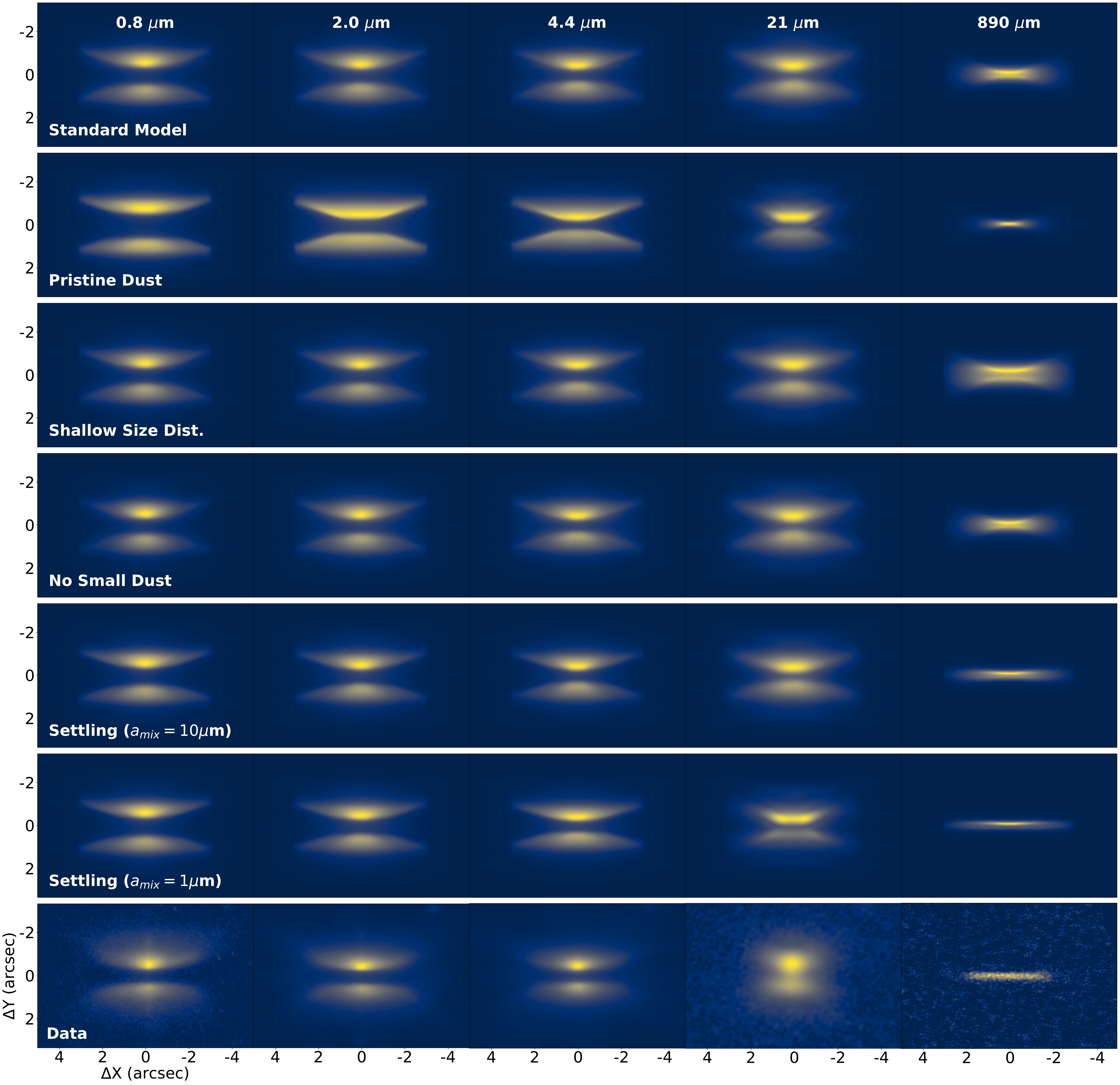}}
    \caption{Model and observed images for various dust grain size distribution and disk structure models. Corresponding parameters are summarized in Table\,\ref{tab:model_res}. All images are shown on a square root stretch except the last column (linear stretch) from 0 to the peak value.}
    \label{fig:model_images_nosettling}
\end{figure*}

\begin{figure*}[htb]
    \centerline{\includegraphics[width=0.49\textwidth]{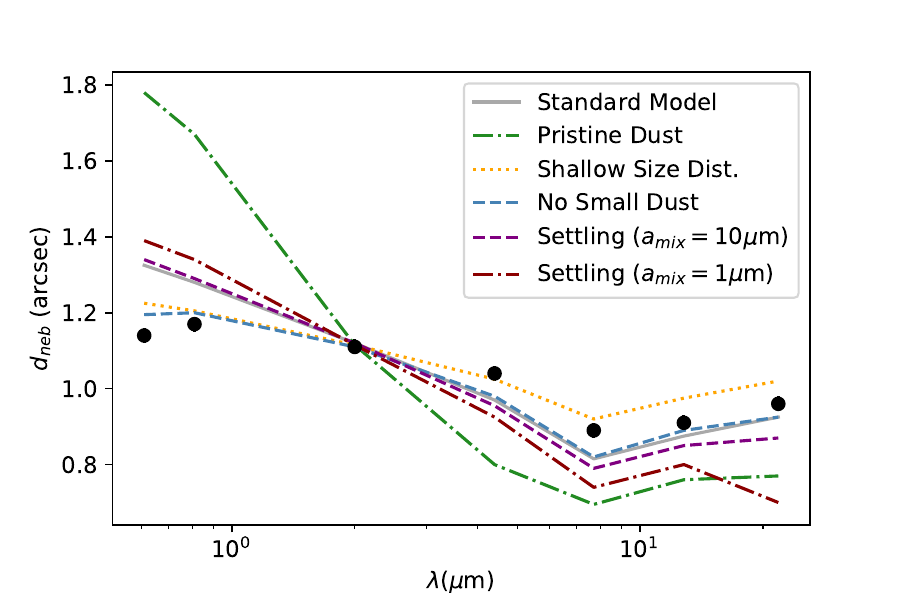}
    \includegraphics[width=0.49\textwidth]{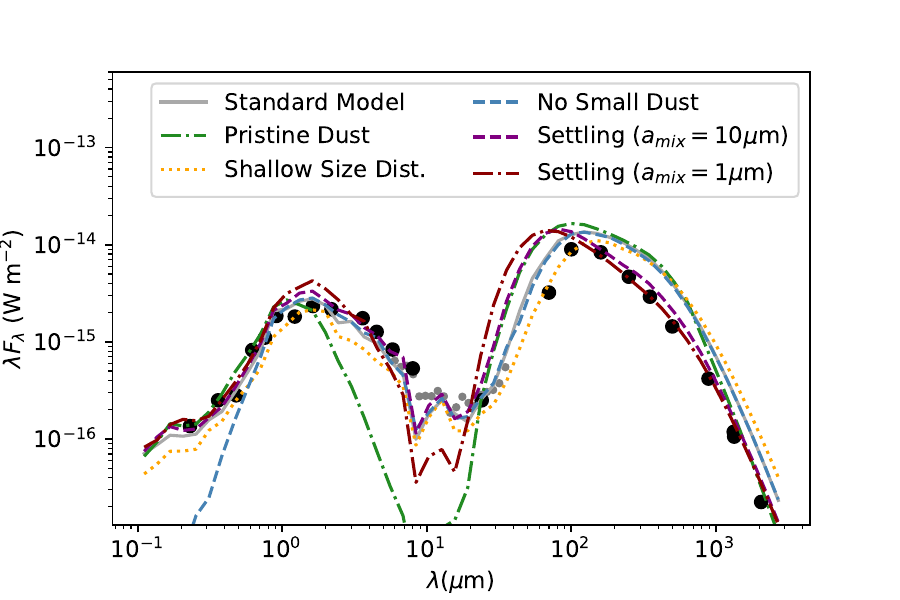}}
    \caption{{\it Left:} Dark lane thickness as a function of wavelength (left panel) for the models explored here (see Table\,\ref{tab:model_res}). {\it Right:} Observed and model SED for the same models. The dip at 9\,$\mu$m in the model SEDs is due to a strong albedo feature from the silicates adopted in the DIANA mixture. We stress that the models are constructed solely on the basis of matching the dark lane thickness at 2\,$\mu$m; all other quantities are merely predicted based on having fit that quantity.}
    \label{fig:model_geometry_nosettling}
\end{figure*}

Figure\,\ref{fig:model_images_nosettling} presents a gallery of PSF-convolved model images from the optical to the sub-mm. Furthermore, Figure\,\ref{fig:model_geometry_nosettling} presents the wavelength dependence of the distance between the top and bottom nebulae (referred to as ``dark lane thickness" in Table\,\ref{tab:model_res}), measured in the same manner as in the data, as well as the observed and model SED. 

Focusing first on the dark lane thickness metric, the standard model displays a decline of about 25\%\ from the optical to the mid-infrared, similar to, albeit slightly steeper, than in the observations. The disk morphology is reasonably well reproduced, although the images are not quite as centrally peaked as in the observations, suggesting that the scattering phase function is not sufficiently forward scattering. The overall SED of the system is also satisfyingly well reproduced, although we note a strong albedo feature around 9\,$\mu$m that is not seen in the {\it Spitzer} observations but is inherent to the dust composition adopted here. The submillimeter regime is where this model shows its largest shortcoming, with the disk much too vertically extended and showing too flared of a structure at large distances from the symmetry axis. This highlights that settling is most likely present in this system. Nonetheless, considering that the model disk mass is only set by the distance between nebulae at 2\,$\mu$m in our modelling approach, we conclude that this model overall matches surprisingly well with all observations.

Conversely, the ``pristine dust" model experiences a strongly chromatic dark lane thickness, with a $\gtrsim50\%$ decline from the optical to the mid-infrared, in line with its much steeper opacity law (Figure\,\ref{fig:dust_prop}). In addition, this model underpredicts the SED by more than an order of magnitude in the 2--20\,$\mu$m range, for lack of effective scatterers in that wavelength regime. Finally, in the submillimeter regime, the model image is much too radially concentrated, as a consequence of the strongly reduced opacity of that dust model at the longest wavelengths. All in all, this model convincingly demonstrates that dust in the \eod\ disk is significantly different from that present in molecular cloud.

To try and reconcile the very shallow chromaticity of the dark lane with a fully mixed model, we considered two minor changes to the standard model. In the first one, we adopted a slightly shallower grain size distribution ($p = 3.25$), while in the other one we increased the minimum grain size to $a_\mathrm{min}=0.3\,\mu$m. Since both of these models produce a flatter extinction law from the optical to the near-infrared (Figure\,\ref{fig:dust_prop}), they succeed in matching the dark lane thickness behavior better than the standard model. However, they still produce submillimeter maps that do not match the observed one. Furthermore, the model with increased minimum grain size underpredicts the system's SED at $\lambda \lesssim0.5\,\mu$m as all dust grains are too large to scatter such short wavelengths. These models indicate that a modest decrease in the population of small ($\lesssim0.1\,\mu$m) relative to the standard model {\bf may} yield a satisfying match to the HST and JWST images, although it would still fail in the ALMA regime.

Finally, we considered three settling models in which grains up to $a_\mathrm{mix}\!=\!100\,\mu$m, $10\,\mu$m or $1\,\mu$m are fully mixed, with large grains increasingly settled with $\eta_\mathrm{settl} = -0.5$. These models yield an effective scale height of $\approx3$, 1 and 0.3\,au at 100\,au for 1\,mm grains, respectively, in reasonable agreement with derived scale height in protoplanetary disks \citep{pinte2016, villenave2022}. The $a_\mathrm{mix}\!=\!100\,\mu$m model is essentially indistinguishable from the ``standard" model in all observations, indicating that this it produces an insufficient degree of settling for mm-emitting grains. For smaller values of $a_\mathrm{mix}$, settled models produce much improved 890\,$\mu$m images, as expected, although the surface brightness along the major axis is not quite flat enough. However, whereas the $a_\mathrm{mix}\!=\!10\,\mu$m model shows a dark lane thickness chromaticity that is almost identical to the standard model presented above, i.e., only slightly too steep, the $a_\mathrm{mix}\!=\!1\,\mu$m model shows a significant decline at the longest mid-infrared wavelengths due to its significant settling of infrared-scattering grains. Furthermore, the SED for that model shows a deeper trough in the mid-infrared whereas the $a_\mathrm{mix}\!=\!10\,\mu$m model matches well with the observed SED, albeit with a modest overprediction in the 30--100\,$\mu$m regime. 

Overall, as summarized in Table\,\ref{tab:model_res}, we consider this brief model exploration as a strong indication that a model in which grains up to $\sim10\,\mu$m in size are fully mixed with the gas and larger grains ($\gtrsim100\,\mu$m) are increasingly settled down to a scale height of $\sim1$\,au could simultaneously match the HST, JWST and ALMA continuum observations of \eod. We note that the disk dust masses required to match the dark lane thickness at 2\,$\mu$m are fairly high, in the $ \sim 3-10 \times 10^{-4}\,M_\odot$ range. This is consistent with the lower limit derived independently by \citet{villenave2020} from ALMA continuum maps. While the dust mass is inversely proportional to the assumed dust opacity and remains therefore somewhat uncertain, the \eod\ disk is on the high end of the observed distribution of $M_d/M_\star$ ratio \citep[e.g.,][]{barenfeld2016}, consistent with it being one the largest disks in Taurus \citep{andrews2018}. We defer additional improvements upon this model, including a different scattering phase function and a more realistic settling prescription, to a further study.

\section{Discussion} \label{sec:disc}

\subsection{Dust properties and settling} \label{subsec:growth_settling}

With a 1000\,au radius as measured in CO, \eod\ is one of the largest EOD in nearby star-forming regions. Its images across all wavelengths are smooth and largely symmetric and it has no known companion. It is therefore an ideal testbed for models of disk structure and dust evolution. The JWST images we presented here allow us to probe the properties of its dust component, mostly opacity and scattering phase function, as well as the degree of vertical settling in the disk.

In scattered light imaging, the thickness of the dark lane that separates the scattering surfaces is driven by the optical depth along our line of sight to the central star. In turn, this depends on the product of the disk mass and the dust opacity. The very modest drop in dark lane thickness from the optical through the mid-infrared indicates that the dust located in the upper layers of the \eod\ disk is characterized by a (nearly) gray opacity law over that wavelength range. Indeed, our analysis showed that a model based on a dust population consistent with that found in molecular cloud is strongly inconsistent with observations. This confirms that significant grain growth occurs in protoplanetary disks. Assuming the standard power law size distribution ($p = 3.5$), the limited dark lane chromaticy in this disk indicates that grains at least as large as 10\,$\mu$m are present in copious amount in its upper layers. Similar conclusions were reached for other disks based on a variety of observation constraints, such as system SED, polarized intensity scattered light imaging or water ice absorption \citep{pontoppidan2007, tazaki2021b, franceschi2023, sturm2023a}. The JWST images presented here provide a direct probe of the dust opacity in the upper outer layers of the \eod\ disk, which complements these other methods. Specifically, the observations presented here demonstrate the abundant presence of 10\,$\mu$m-sized grains in a region located $\sim60$\,au above the midplane at a radius of $\sim400$\,au from the central star. Ultimately, combining these different observables, as well as the spectral shape of the mid-infrared silicate features, will enable a more granular analysis of the dust properties in the disk upper layers, including allowing for departures from simple power law parametrizations. Indeed, each observational approach is uniquely sensitive to different aspects of the dust component, both spatially and in terms of grain size.

One of the shortcomings of our ``standard" model, however, is that it predicts a change in dark lane thickness between the optical and near-infrared ranges that is 10--15\%\ too large. An even shallower dust opacity law seems necessary to account for this, as supported by our models using a shallower size distribution or a larger minimum grain size. This is an indication that the growth of dust particles may introduce a break in the slope of the distribution, as is suggested by numerical models of grain growth and fragmentation \citep[e.g.,][]{birnstiel2011}. However, it is unclear if grains as large as 10\,$\mu$m can efficiently grow in situ in the uppermost layers of the outer disks, where overall densities are relatively low. Instead, a (turbulent) mechanism may be necessary to loft them up from deeper layers, where they could more easily form \citep[e.g.,][]{tazaki2021b, franceschi2023}.

While the fully mixed models we developed can (roughly) reproduce all HST and JWST images of \eod, they produce submillimeter images that do not match the observations. In short, the optical depth through the disk is much too high, and thermally emitted photons scatter off layers that are a few tens of au above the midplane. In our models, we were able to obtain a better match to observations by introducing some settling of large grains, as suggested by other ALMA observations of protoplanetary disks. While we did not consider physical models of dust settling, our parametric approach allows us to reach some general conclusions. To reproduce the images of \eod\ through the HST--JWST regime, we find that grains up to $\gtrsim10\,\mu$m must be vertically fully mixed, i.e., turbulence is sufficiently strong to dredge them all the way to the disk surface. Conversely, millimeter-emitting grains (sizes $\gtrsim100\mu$m) must have a scale height that is roughly one tenth that of the gas and the smallest dust grains. 

Let us now consider how this compares to settling model predictions. Pure hydrodynamical simulations \citep[e.g.,][]{dubrulle1995} indicate that the scale height of the dust component scales as $h_\mathrm{dust} \approx h_\mathrm{gas} \sqrt{\frac{\alpha}{\alpha + St}}$, where $St$ is the Stokes number, i.e. the product of the orbital frequency by the stopping time of a grain as a result of the drag force induced by the surrounding gas, and $\alpha$ the standard turbulence parameter \citep{shakura1973}. In the large grain limit, $St \propto a$, leading to $h_\mathrm{dust} \propto a^{-0.5}$ in the case of weak turbulence ($\alpha \ll St$). Therefore, mm-sized dust are expected to have a scale height that is $\approx 10$ times smaller than that of 10\,$\mu$m grains, in line with our findings. Magneto-hydrodynamical models, on the other hand, predict shallower dependencies: \cite{fromang2009} find an $h_\mathrm{dust} \propto St^{-0.2}$ approximate relationship, for instance. This may imply that the importance of magnetized transport is reduced in the outer regions of the \eod\ disk, which are primarily traced by scattered light images. However, \cite{riols2018} have shown that, while the hydrodynamical solutions may be approximately correct in the context of 1D models, higher dimension simulations reveal more complex behaviors, in which it is nearly impossible to characterize settling through a single scale height quantity. To make matters worse, both types of simulations concur on the fact that the relevant quantity is not directly the size of dust grains but their Stokes number. Because the latter depends also on the local gas density and the material density of the dust particles, the constraint we have obtained from the JWST images presented here can only be indirectly compared to such models. We leave the comparison of more physically-grounded settling models with JWST images to future studies.

While the models we have constructed reproduce the dark lane thickness, they are imperfect when considering the overall morphology of the \eod\ disk. In particular, model images appear consistently less centrally peaked than is observed, especially at the longest JWST wavelengths. Although this may be due to a combination of factors, we suspect that a main issue is that the dust prescription we have used leads to a scattering phase function that is not sufficiently forward scattering. The more peaky phase functions could be either due to solid grains with a lower refractive index (i.e., more ice rich) or porous aggregates. Indeed, lower refractive indices of solid grains facilitate efficient transmission of light, leading to more efficient forward scattering \citep[e.g., Figures 10 and 17 in][]{tazaki2021}.  Porous aggregates are also known as efficient forward scatterers \citep{min2012, tazaki2019}, although the phase function depends on the detailed structure of aggregates, such as fractal dimension \citep{min2016, tazaki2016, tobon2022}. Future modeling of this system should explore such dust properties to improve the match between observed and model scattered light images of \eod\ (R. Tazaki, priv. comm.).

Finally, we note that the modest chromaticity of the dark lane thickness in \eod\ is in stark contrast with previously-imaged EODs, which become markedly thinner towards longer wavelengths, at least out to 5\,$\mu$m \citep[][see also Figure\,\ref{fig:neb_sep}]{duchene2010, mccabe2011}. One possible explanation is that this system contains a shallower size distribution than other disks, possibly as a result of more advanced dust evolution. It could also be the case that the \eod\ disk is more optically thick than other EODs. Indeed, in the limit of very high optical depths, the chromatic dependence of the opacity law is gradually muted with increasing disk mass as the elevation of $\tau = 1$ surface does not depend much of the total column density in the disk since the density drops rapidly in the disk atmosphere. In other words, if the disk is very optically thick up to $\approx60au$ above the midplane at $\lambda<21\,\mu$m and the density drops extremely fast above this layer, the lack of chromaticity could be explained with a wide range of grain sizes. The gradual decline in surface brightness in the optical HST images seems to argue against this scenario but further analyses are needed to definitely rule it out. JWST observations of other EODs will allow us to place \eod\ in the larger context. 

\subsection{A possible disk wind} \label{subsec:wind_pahs}

Arguably the most surprising outcome of the JWST observations of \eod\ is the X-shaped feature that is prominent at 7.7 and 12.8\,$\mu$m and resides immediately above the warm molecular layer traced by the ALMA $^{12}$CO mapping, at elevations up to $\sim225$\,au. Without follow-up spectroscopic observations, no definitive conclusion can be reached about its nature. Given its large opening angle ($\sim35$\degr\ above the midplane), this feature must arise in the highest layers of the disk, or in a wind above the disk, and over a broad range of stellocentric distances. In this context, it is worth recalling two other unusual features of this system: 1) the presence of scattered light in the optical up to $\gtrsim250$\,au above the midplane, and 2) the unusually extended CO channel map at a very small velocity relative to the systemic velocity, also marking surfaces that are about $\sim325$\,au above the midplane. The similarity in elevation between these various features suggest a common origin.

Since the diffuse optical emission observed in this region is found in two separate filters, it is unlikely to be due to line emission and instead points to a scattered light nature. The absence of this component in even the shortest wavelength JWST image suggests scattering off of small ($\lesssim1\,\mu$m) grains. Conversely, the X-shaped feature is unlikely to be solely due to scattering since it is not detected in the 4.4\,$\mu$m image. We speculate that it could be due to H$_2$ photodissociation, or to out-of-equilibrium thermal emission from very small ($\lesssim0.1\mu$m) grains in the disk atmosphere that are in direct view of the X- and UV-emitting central source. In the H$_2$ interpretation, we note the lack of detection of the S(0) line (at 28.2\,$\mu$m) in the IRS spectrum and the absence of the X-shaped feature in the 4.4\,$\mu$m image, which contains the S(9) line (at 4.7\,$\mu$m). In conjunction, these would suggest an excitation temperature of $\lesssim100$\,K \citep{nomura2005, lahuis2007}. Polycyclic aromatic hydrocarbons (PAHs) are commonly observed in the 3--15\,$\mu$m range in disks surrounding more massive Herbig AeBe stars \citep{meeus2001}, and it is plausible that the active accretion from the central source in \eod\ can excite such grains even though it is not intrinsically as hot \citep{siebenmorgen2012}. We note that, while the X-shaped feature is prominent in the JWST images presented here, it represents a small fraction of the integrated brightness of the system, so that PAHs could remain undetected in the {\it Spitzer}/IRS spectrum. It is also possible that the out-of-equilibrium grains have a more standard composition, in which case their stochastic emission would appears as a sum of warmer-than-expected blackbodies \citep{draine2001}. Ultimately, the only requirement for this process to occur is that the grains are very small, to the point where the absorption of a single UV photon can overheat them for a short time. If this very small grain interpretation is correct, it could be common among T\,Tauri systems, and mid-infrared integral field unit spectroscopy of a large sample could confirm this conclusion.

A plausible scenario to explain these various features invokes a photodissociative disk wind \citep[e.g.,][]{owen2011, bai2016}. In this picture, the UV emission from the central star (and accretion column) dissociates molecules located at the very top of the warm molecular layer. The excited atoms escape the disk surface and are then placed on ballistic outward trajectories. Recombination of H$_2$ molecules along them could then produce the X-shaped feature through limb brightening of the conical wind, while the non-Keplerian flow of the wind could possible lead to the odd-looking low-velocity CO channel map. It is also possible that PAHs and other very small grains are entrained along with the gas and get stochastically heated once they reach high enough layers from which they are in direct sight of the central source. Furthermore, if this wind can entrain grains as large as 0.1--1\,$\mu$m, this would naturally produce a ``veil" of optical scattered light well above the disk midplane, as is observed in \eod. Current disk wind models suggest that entrainment of such grains is possible \citep[e.g.,][]{booth2021, rodenkirch2022, franz2022}. In summary, such a scenario can qualitatively reproduce some of the unexpected features observed in this system, but we defer to a later study a physically-grounded wind model.

\section{Conclusion} \label{sec:concl}

In this study, we have presented new optical through suubmillimeter high-resolution observations of the \eod\ edge-on protoplanetary disk. With a $\approx1000$\,au radius, this disk is one of the largest disks known in nearby star-forming regions, and it shows a smooth and symmetrical structure. This makes it a perfect prototype to study the disk structure and dust properties in such environments.

We have presented the first JWST observations of \eod, imaging its edge-on protoplanetary disk from 2 to 21\,$\mu$m, which we interpret in combination with HST images at 0.6 and 0.8\,$\mu$m. The disk is observed in scattered light throughout the entire wavelength range considered here, and shows very modest levels of chromaticity. In particular, the thickness of the dark lane that separate the two disk surfaces only declines by $\sim15$\% from 0.6 to 21\,$\mu$m, a much shallower decline than has been observed for other edge-on disks (over more limited wavelengths ranges). We also find that up to 21\,$\mu$m, the vertical extent of the disk is an order of magnitude larger than observed in ALMA continuum maps, a confirmation that the large, millimeter-emitting grains, must be strongly settled. To interpret these observations, we build radiative transfer models that explore the role of the grain size distribution and degree of dust settling in the disk. We conclude that grains up to $\gtrsim10\mu$m in size are fully coupled to the gas up to the disk surface layers, whereas  particles larger than $\gtrsim100\mu$m must be concentrated $\sim10$ times closer to the midplane. This is the first time observations provide a direct constraint on the degree of settling of grains in the 10\,$\mu$m size regime via a spatially-resolved analysis of dust opacity. Future studies of other edge-on disks will help place this system in context and identify the underlying physical processes that are responsible for settling and, ultimately, growth, of dust particles in disks.

In addition to analyzing the scattered light properties of the disk, the HST and JWST images reveal multiple knots of emission from the jet launched by the central source. Multiple emission line features are also detected in the {\it Spitzer}/IRS spectrum of the system. The ALMA CO observations yield an estimate of the stellar mass ($\sim0.4\,M_\odot$) that is on the low end of the range suggested by past spectroscopic observations. Finally, we have also identified several unexpected features in this system. The HST optical images point to the presence of $\lesssim1\,\mu$m grains located up to 250\,au above the disk midplane. The $^{12}$CO channel maps reveal a $\sim300$\,au-elevation CO layer with unusual kinematic properties, suggesting non-Keplerian motion. Finally, and most intriguingly, the 7.7 and 12.8\,$\mu$m JWST images reveal a X-shaped feature that is located above the CO warm molecular layer. We propose that these various features can be explained if the disk drives a significant photoevaporative disk wind that entrains small dust grains. Follow-up spectroscopic observations of this system can help identify the nature of the X-shaped feature and its relationship to the other phenomena.


\begin{acknowledgments}
This project has received funding from the European Research Council (ERC) under the European Union's Horizon Europe research and innovation program (grant agreement No. 101053020, project Dust2Planets, PI F.M\'enard). GD, KRS, MV, and SGW acknowledge funding support from JWST GO program \#2562 provided by NASA through a grant from the Space Telescope Science Institute, which is operated by the Association of Universities for Research in Astronomy, Incorporated, under NASA contract NAS5-26555.  We are grateful to the W.M. Keck Institute for Space Studies as the hosts for our week-long team meeting in May 2023. CP acknowledges Australian Research Council funding  via FT170100040, DP18010423 and DP220103767. RT acknowledges financial support from a CNES postdoc fellowship. MV research was supported by an appointment to the NASA Postdoctoral Program at the Jet Propulsion Laboratory, administered by Oak Ridge Associated Universities under contract with NASA. Some of the data presented in this paper were obtained from the Mikulski Archive for Space Telescopes (MAST) at the Space Telescope Science Institute. The specific JWST observations analyzed can be accessed via \dataset[doi:10.17909/rgg7-zk46]{https://doi.org/10.17909/rgg7-zk46}. This paper makes use of the following ALMA data: ADS/JAO.ALMA\#2016.1.0771.S. ALMA is a partnership of ESO (representing its member states), NSF (USA) and NINS (Japan), together with NRC (Canada), MOST and ASIAA (Taiwan), and KASI (Republic of Korea), in cooperation with the Republic of Chile. The Joint ALMA Observatory is operated by ESO, AUI/NRAO and NAOJ. The National Radio Astronomy Observatory is a facility of the National Science Foundation operated under cooperative agreement by Associated Universities, Inc.
\end{acknowledgments}

%

\vspace{5mm}
\facilities{JWST (NIRCam and MIRI), HST (ACS), ALMA}






\appendix

\section{Compiling the SED of the system \label{sec:sed}}

To produce the complete SED of the system, we complement the new JWST photometric datapoints presented in this study and the IRS spectrum (Section\,\ref{subsec:dustdisk}) with published and archival data. The complete SED is presented in Figure\,\ref{fig:sed} while Table\,\ref{tab:sed} summarizes the photometry adopted here (see below), as well as the new JWST fluxes. 

\begin{deluxetable}{cccc}
 \tablecaption{Adopted system SED for \eod \label{tab:sed}}

 \tablehead{
 \colhead{$\lambda$ ($\mu$m}) & \colhead{$F_\nu$ (mJy)} & \colhead{Observatory} & \colhead{Reference}}
 \startdata 
 0.23 & 0.010 & GALEX & 1 \\
 0.36 & 0.030 & SDSS & 2 \\
 0.48 & 0.045 & '' & 2 \\
 0.62 & 0.17 & '' & 2 \\
 0.76 & 0.28 & '' & 2 \\
 0.91 & 0.56 & '' & 2 \\
 1.22 & 0.74 & 2MASS & 3 \\
 1.63 & 1.27 & '' & 3 \\
 2.19 & 1.58 & '' & 3 \\
 3.6 & 2.1 & {\it Spitzer}/IRAC & 4 \\
 4.5 & 1.9 & '' & 4 \\
 5.8 & 1.61 & '' & 4 \\
 8.0 & 1.42 & '' & 4 \\
 24 & 2.0 & {\it Spitzer}/MIPS & 4 \\
 70 & 75 & '' & 4 \\
 100 & 298 & {\it Herschel}/PACS & 5 \\
 160 & 445 & '' & 5 \\
 250 & 394 & {\it Herschel}/SPIRE & 6 \\
 350 & 340 & '' & 6 \\
 500 & 236 & '' & 6 \\
 890 & 124 & ALMA & 7 \\
 1330 & 52 & SMA & 8 \\
 1340 & 47 & ALMA & 7 \\
 2060 & 15 & '' & 7 \\
 \hline
 \multicolumn{4}{c}{New JWST Photometry}\\
 \hline
 2.0 & 2.4 & JWST/NIRCam & 6 \\
 4.4 & 2.6 & JWST/NIRCam & 6 \\
 7.7 & 2.3 & JWST/MIRI & 6 \\
 12.8 & 1.8 & JWST/MIRI & 6 \\
 21.0 & 1.9 & JWST/MIRI & 6 \\
 \enddata
\tablecomments{References: 1) GALEX GR6/GR7; 2) \citet{ahumada2020}; 3) \citet{cutri2003}; 4) \citet{rebull2010}; 5) \citet{herschel2017}; 6) This work; 7) \citet{villenave2020}; 8) \citet{andrews2013}.}
\end{deluxetable}

\eod\ was detected by GALEX with its NUV channel (0.18--0.28\,$\mu$m) in a 2662\,s exposure as part of the Medium Imaging Survey. We retrieved the corresponding flux density from the GR6/GR7 source catalog available through the MAST portal\footnote{{\tt https://galex.stsci.edu/GR6/?page=mastform}}. The GALEX detection suggests that the source is actively accreting, consistent with the presence of the collimated atomic jet, and that it is not significantly extincted by foreground cloud material. No corresponding FUV (0.13--0.18\,$\mu$m) observation was taken. In the optical range, we adopted the SDSS DR16 \citep{ahumada2020} ``Petrossian" magnitudes of \eod. These represent an aperture photometry method with a radius based on the extent of the source; while this method was designed for galaxies, we assume that it operates reasonably well for any smooth, extended source. In the far-infrared, \eod\ was observed with {\it Herschel} as part of the ``{\it Herschel} Gould Belt Survey" Guaranteed Time Key Program \citep{andre2010}. PACS flux densities are available through the {\it Herschel} PACS Point Source Catalog \citep{herschel2017}. To complete the SED, we retrieved the level 2.5 SPIRE (250, 350 and 500\,$\mu$m) data products from the {\it Herschel} archive and averaged the individual frames to produce final images in each filter. \eod\ is clearly detected -- and unresolved -- in all filters. Following the SPIRE Handbook, we estimated photometry for the system using Gaussian fitting and applying band-appropriate aperture corrections to obtain the far-infrared flux densities of \eod. The resulting measurements, for which we estimate a 10\% relative uncertainty due to a combination of absolute calibration error and uncertainties due to the complex background, are listed in Table\,\ref{tab:sed}. We note that a ``starless core" was identified by \citet{marsh2016} at the location of \eod, using the same {\it Herschel} data. The SPIRE flux densities estimated by these authors are broadly consistent with our estimates. However, their 160\,$\mu$m measurement is about 33\% lower than that estimated by \citet{herschel2017}, probably as a result of methodological differences. As a result, we decided against adopting their {\it Herschel} photometry both to ensure consistency and to avoid an implausible break between 100 and 160\,$\mu$m.

In some regimes, \eod\ is detected in multiple catalogs and surveys, which reveals significant variability in some cases.  In the optical, the Pan-STARRS Kron photometry \citep{chambers2016} is consitent with the SDSS photometry adopted here to within 20\%\ or so. On the other hand, the UKIDSS photometry \citep{lawrence2012} is a factor of $\approx2$ fainter than the 2MASS photometry, and the 3.4 and 4.6\,$\mu$m WISE flux densities \citep{cutri2012} are about 30\% lower than the corresponding Spitzer/IRAC ones \citep{rebull2010}. Finally, the integrated flux densities measured in the JWST images presented here are 20--60\% higher than the {\it Spitzer} measurements in the 4--8\,$\mu$m, whereas the 20\,$\mu$m JWST flux is consistent with the 24\,$\mu$m {\it Spitzer} one. All in all, the source is affected by a typical variability of up to a factor of $\approx2$ from the optical to the mid-infrared, which is in line with the morphological variability noted by \citet{luhman2009} in the near-infrared. Here we selected datasets on the basis of producing a smoothly continuous SED but we caution against interpreting any photometric point without taking into consideration this significant variability.

Finally, we performed power law fits to the (sub)millimeter and UV/blue regimes of the SED, which led to spectral indices of $\alpha_\mathrm{mm} = 2.5$ \citep[see also][]{villenave2020} and $\alpha_\mathrm{UV} = -2.0$, defined by $F_\nu \propto \nu^\alpha$. In particular, at $\lambda \lesssim0.5\,\mu$m, the system's SED presents a markedly different spectral index from the rest of the optical regime, suggesting the presence of a significant accretion-driven UV excess and we use the latter index in our modeling setup (Section\,\ref{subsec:models_setup}).


\section{Dust opactiy in molecular clouds \label{sec:dust_molcloud}}

The extinction law in the interstellar medium (ISM) and molecular clouds has long been used as a probe into the grain size distribution in these environments \citep[][and references therein]{draine2003}. In particular, the realization that the extinction law in molecular cloud flattens out in the near- to mid-infrared range revealed the presence of larger grains in these clouds than in the diffuse ISM \citep[e.g.,][]{lutz1999, flaherty2007}, suggesting that the initial stages of grain growth occur before the formation of stars and their disks. To determine the extent and timescale of grain growth in disks, it is therefore necessary to develop a model of the dust population of molecular clouds, whose properties can then be compared to observations of protoplanetary disks, such as those presented in the present study.

Unlike the case of the diffuse Galactic ISM, the extinction law in molecular clouds is not a unique curve, as it depends, among other things, on the temperature and column density of the cloud. Furthermore, no single observational method can probe extinction from the short-wavelength optical to the mid-infrared, requiring combining multiple studies that may inherently probe different environments. Fortunately, the agreement between studies is sufficient for our purposes, as we show below.

\begin{figure}[htb]
    \centerline{\includegraphics[width=0.5\columnwidth]{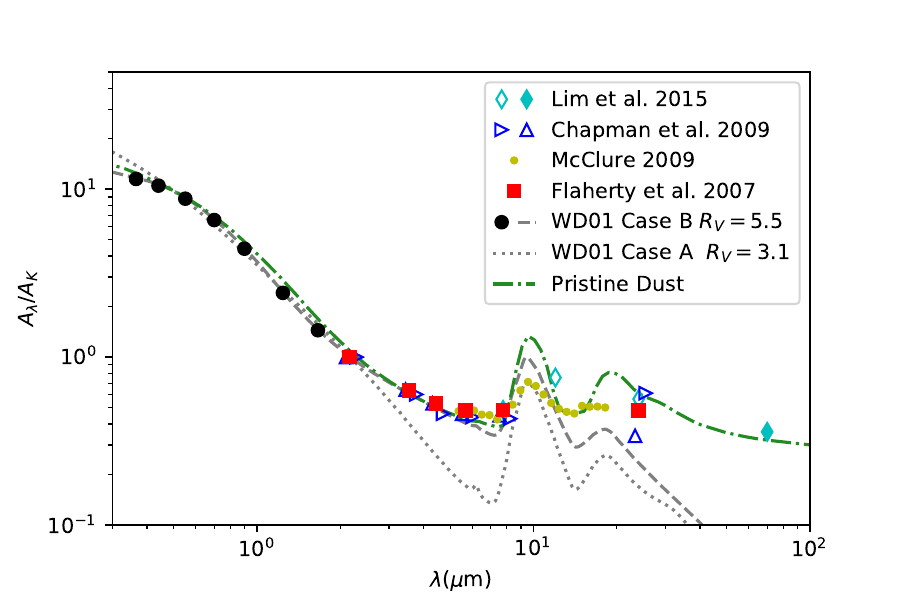}}
    \caption{Near- to far-infrared extinction law measured in molecular clouds (colored symbols) as well as function forms for the interstellar extinction from \cite{weingartner2001} (labeled WD01 in the figure). The Case A and B curves are appropriate for the diffuse ISM and molecular clouds, respectively. Solid circles represent Case B extinction values for standard optical and near-infrared filters. The two sets of blue triangles represent the $1 \leq A_K < 2$ and $A_K \geq 2$ results from \citet{chapman2009}; the latter corresponds to the lowest value of $A_{24}/A_K$. The dotted orange line indicates the opacity law of the ``pristine dust" model that was constructed to match the adopted molecular cloud extinction law (represented by the solid symbols).}
    \label{fig:ext_molcloud}
\end{figure}

To construct the extinction law in the molecular clouds, we focus on {\it Spitzer} studies. Specifically, \cite{flaherty2007} and \cite{chapman2009} provided broadband extinction laws ranging from 3.6 to 24\,$\mu$m for multiple nearby star-forming regions. In both cases, we computed the weighted average of the individual line of sight extinction laws. We note that the $1 \leq A_K < 2$ and $A_K \geq 2$ curves from \cite{chapman2009} bracket the 24\,$\mu$m estimate from \cite{flaherty2007} and are comparable to the latter at $\lambda \leq 8\,\mu$m. \cite{lim2015} extend broadband extinction through an infrared dark cloud to longer wavelengths (12, 24 and 70\,$\mu$m). Again, we computed the average of the various line of sights they studied and anchored their relationship to the 8\,$\mu$m extinction coefficient from \cite{flaherty2007}. Finally, \cite{mcclure2009} used 5--20\,$\mu$m spectroscopy to sample the extinction law across the silicate feature. To improve signal-to-noise, we resampled their extinction curve to a resolution of $R\approx15$. Altogether, we find that these different extinction laws agree to $\approx10\%$ across the near- to mid-infrared range (Figure\,\ref{fig:ext_molcloud}). To produce a representative extinction law, we adopt the \cite{flaherty2007} datapoints from 3.6 to 24\,$\mu$m, together with the 70\,$\mu$m point from \cite{lim2015} and the resampled IRS spectrum from \cite{mcclure2009}. At shorter wavelengths, we adopted the "Case B" $R_V=5.5$ curve from \cite{weingartner2001}, as it has been found to match extinction in molecular clouds out to at least 5\,$\mu$m. We emphasize that this is an hybrid extinction law that we only use as a general guideline.

We then constructed a ``pristine dust" model that attempts to match the observed extinction law in molecular clouds. We assume the same composition as in our dust models (see Section\,\ref{subsec:models_setup}). No single power law size distribution can simultaneously reproduce the steep drop-off across the optical and near-infrared and the nearly flat extinction law across the mid-infrared. We therefore used a model that combines two power laws: a $p=-3$ power law for grains sizes ranging from 0.03\,$\mu$m to 0.35\,$\mu$m, and a $p=-1.5$ power law from 0.35\,$\mu$m to 100\,$\mu$m. The normalization constants for both power laws are such that 80\% of the total mass is in the large grains component. This model produces an extinction law that matches well (within $\approx$10\%) all broadband estimates of the extinction law, although it produces mid-infrared silicate features that are too strong (Figure\,\ref{fig:dust_prop}). This is likely a result of the silicate-rich nature of the adopted composition coupled with the large contribution of the smallest dust grains. However, any attempt to suppress the silicate feature in our simplified framework led to a much poorer match to the broadband extinction law. This indicates that this dust model is not physically accurate, which is further reinforced by the sharp break at a grain size of 0.35\,$\mu$m. Nonetheless, this model produces the correct broadband opacity across the broad range of wavelengths of interest for our study and thus we consider that it is a good proxy for molecular cloud dust.


\bibliography{gduchene}{}
\bibliographystyle{aasjournal}



\end{document}